\documentstyle[twocolumn,prb,aps]{revtex}

\tightenlines
\begin{document}
\twocolumn[
\hsize\textwidth\columnwidth\hsize\csname@twocolumnfalse\endcsname

\draft

\title{Electronic Raman Scattering and Phonon Self-Energy Effects  
          in the RE-123  System:
          Signatures of Gap and Pseudogap}

\author{Andreas Bock}

\address{Institut f\"ur Angewandte Physik und Zentrum f\"ur 
          Mikrostrukturforschung, \\ 
          Universit\"at Hamburg, Jungiusstra{\ss}e 11, D-20355 Hamburg, 
          Germany}

\date{\today}

\maketitle

\begin{abstract}

Using Raman spectroscopy and transport measurements 
we investigate thin epitaxial films of 
$\rm Y_{1-x}(Pr,Ca)_xBa_2Cu_3O_{6+y}$.
We explore the electronic Raman 
responses in $\rm A_{1g}$, $\rm B_{1g}$, and $\rm B_{2g}$ symmetry 
obtained after subtraction of phononic excitations, 
and especially, the $2\Delta$ peaks that form out of the electronic 
background below $T_c$.
We find that the energy of the $\rm B_{1g}$ $2\Delta$ peak,
which is a measure of the gap value, increases 
monotonically with decreasing doping until the peaks become unresolvable. 
In contrast, the peaks in $\rm A_{1g}$ symmetry follow $T_c$ being still 
resolvable in the Pr-doped films.
The $\rm B_{2g}$ responses are weak and a $2\Delta$ peak is only detected 
at the highest doping level.
As a consequence of strong electron-phonon coupling, the $\rm B_{1g}$ 
phonon at $\sim 340$ cm$^{-1}$ exhibits a pronounced Fano-type 
line shape. 
We use a phenomenological model to describe the line shape that takes 
into account real and imaginary part of the electronic response.
It allows us to obtain the self-energy corrections and the 
mass-enhancement factor $\lambda$ as a measure of the coupling.
In the normal state we find $\lambda=0.015$ around optimal doping and 
decreasing values with decreasing doping.
The electron-phonon coupling increases strongly below 
$T_c$ in overdoped samples in which the $\rm B_{1g}$ $2\Delta$ peaks 
appear in the vicinity of the phonon.
Self-energy effects observed in the superconducting 
state can only partly be assigned to the redistributing electronic 
response.
Anomalies with respect to frequency, linewidth, and intensity remain.
They appear at increasing temperatures with decreasing doping
and we provide evidence that they are connected to the presence 
of the pseudogap.
We supplement our study by a comparison with single crystal data 
and investigate the influence of site-substitutional disorder on the 
electronic response and the electron-phonon interaction.

\end{abstract}


]  

\narrowtext

\section{introduction}

Since the discovery of the cuprate high-temperature 
superconductors (HTS)\cite{Bednorz86} 
it has been established that the important common feature of all 
cuprates are the copper-oxygen (CuO$_2$) planes.
Even though not discovered yet, it is believed that the mechanism 
causing the high transition temperatures $T_c$ is 
located in and related to these planes.
Often discussed candidates for the pairing mechanism are an exchange of 
antiferromagnetic fluctuations\cite{Scalapino,Pines} or interlayer 
tunneling.\cite{Anderson,Lee}
The role phonons play in the mechanism is not 
resolved.\cite{Krakauer93,Peter98,Hadjiev98b}
Isotope experiments indicate that phonons may be 
involved,\cite{Franck}
at least away from optimal doping which is defined by 
$\sim 0.16$ holes per plane in the unit cell.
However, the observed\cite{Franck,Friedl90}
and calculated \cite{Krakauer93,Rodriguez90}
electron-phonon couplings are too small to explain $T_c$'s above 
$\sim 50$ K.
Nevertheless, it is noted that the electron-phonon coupling in 
the cuprate HTS is another example of their exotic 
properties.\cite{Migdal}

Difficult to understand in the framework of phonon-mediated pairing 
are the various peculiarities of the superconducting gap that have been 
established so far.
The most important one concerns its symmetry which appears to be 
predominantly of $d_{x^2-y^2}$-wave 
type.\cite{vanHarlingen,Kirtley96,Shen93} 
Others are the strange dependencies of the gap on temperature and on 
doping observed with various spectroscopic methods like e.g. tunneling, 
angle-resolved photoemission (ARPES), or Raman spectroscopy.
Moreover, it is found that a pseudogap exists in the 
excitation spectra of underdoped cuprates 
above $T_c$.\cite{Warren89,Loram93,Homes93}
Recent ARPES studies indicated that gap and pseudogap 
are intimately related.
In detail, it was observed that in underdoped compounds a gap opens 
up at the Fermi surface (FS) crossings 
along the $(\frac{\pi}{a},0)$ to $(\frac{\pi}{a},\frac{\pi}{a})$ directions 
at the pseudogap temperature $T^*$.\cite{Harris96,Norman98}
Below that temperature the gapped regions grow continuously, 
covering the entire FS at $T_c$.\cite{Norman98} 
However, recent scanning tunneling microscopy 
measurements,\cite{Renner98a} indicated an increasing pseudogap 
above $T_c$ which disappeared close to room temperature independent 
of doping.
An intimate relation of both gaps is predicted in the precursor scenario as 
presented by Emery and Kivelson\cite{Emery95}
in which preformed pairs exist above the transition temperature but no 
phase coherence between the pairs is established.
Both gaps are assumed to be unrelated in scenarios in which the 
pseudogap is associated with competing magnetic or charge density 
wave instabilities.\cite{UnrelPseudogap}

Even though $T_c$ exhibits a maximum at optimal doping,
various spectroscopies revealed that the gap near 
$T=0$ increases in a monotonic 
fashion with decreasing doping.\cite{White96,Miyakawa98,Chen93} 
This behavior contradicts with mean-field pictures according to which the 
ratio $2\Delta(T=0)/\mathrm{k}_{B}$$T_{c}$ should be constant.
The behavior is, however, in agreement with pictures in which $T_c$
in the underdoped regime
is regarded as the temperature of a Bose-Einstein condensation of 
preformed pairs.\cite{Emery95,Uemura89} 
In the latter case the increasing gap energy with decreasing doping 
can be related to the concomitantly increasing strength of 
the antiferromagnetic correlations\cite{Blumberg97} 
and the decreasing $T_c$ 
to the decreasing density of cooper pairs.\cite{Uemura89} 
Whereas gap features are observed in tunneling and ARPES 
spectroscopy at all dopings, the $2\Delta$ peaks
measured in Raman spectroscopy diminish strongly in the underdoped 
region of the phase 
diagram.\cite{Blumberg97,Kendziora95,Chen97,Nemetschek97} 
The temperature dependence of the $2\Delta$ peaks is 
weak,\cite{Boekholt91,Gasparov99} and commonly they vanish
above $T_c$.
However, recently a small doping-independent feature has 
been reported to persist into the normal state in underdoped 
$\rm Bi_2Sr_2CaCu_2O_{8-\delta}$ (Bi-2212).\cite{Blumberg97,Quilty98}.
Indications of the pseudogap in electronic Raman scattering 
of underdoped 
cuprate HTS so far, are reductions of spectral weight seen in 
$\rm B_{2g}$ (Ref. \onlinecite{Nemetschek97}) and 
$\rm B_{1g}$ symmetry.\cite{Blumberg97,Chen97} 
In addition, anomalies have been observed around $T^*$ 
with respect to phonon frequencies\cite{Litvinchuk92} 
and intensities of interband electronic 
excitations.\cite{Ruani97} 
Whereas pseudogap features above $T_c$ are consistently observed 
by various techniques in 
the underdoped region of the phase diagram it is 
currently under debate whether a pseudogap is also present in 
overdoped compounds.\cite{Renner98a,Tallon98,Gupta98}

Using Raman spectroscopy not only electronic but also 
phononic excitations can be investigated.
In this regard, the renormalization of phonons due to the interaction 
with the pair-breaking excitations is of particular interest.
A famous example for these kind of studies is 
the $\rm B_{1g}$ Raman-active phonon at $\sim 340$ cm$^{-1}$ in 
$\rm REBa_2Cu_3O_{7-\delta}$ (RE-123) with RE a rare-earth atom or 
Y.\cite{Macfarlane87,Cooper90b}
This phonon represents an out-of phase vibration of the planar 
oxygens\cite{Thomsen91Rev} and couples remarkably strong to electronic 
excitations.
The strength of the coupling is related to the buckling of the 
CuO$_2$ planes in RE-123 compounds,\cite{Thomsen90buck} 
and the $\rm B_{1g}$ 
phonon is much weaker in compounds in which the planes are flat, e.g. in 
Bi-2212.\cite{Devereaux98}
Moreover, it was shown that the coupling is amplified by 
the crystal field originating from the different valences 
at the RE and the Ba site on both sides of the plane.\cite{Devereaux95Pho}
As the symmetry of the $\rm B_{1g}$ 
phonon matches that of the order parameter and as its energy is 
close to $2\Delta$, at least for overdoped RE-123, 
the $\rm B_{1g}$ 
phonon turns out to be particularly well-suited for the investigation 
of the electron-phonon coupling. 
Neutron experiments have shown that the strongest coupling of this 
phonon is observed for $q$ along the $[100]$-direction where the 
$d_{x^2-y^2}$-wave order parameter has its maximum 
size.\cite{Reznik95}
There, the most pronounced self-energy effects are observed in the vicinity 
of the Brillouin zone 
center which is probed by Raman spectroscopy.

Due to its coupling to low-energy electronic excitation 
the $\rm B_{1g}$ phonon has a Fano-type line shape. 
The change of the 
line-shape parameters at low temperatures was therefore identified as 
originating from a superconductivity-induced redistribution of 
the electronic excitation spectrum.\cite{Cooper90b}
The renormalization of this phonon in various RE-123 polycrystals, in 
which 
either the rare-earth atom or the oxygen isotope or both have been 
substituted in order to shift the phonon frequency, 
could be accounted for 
in a strong-coupling theory by Zeyher and Zwicknagel\cite{Zeyher90} 
using an $s$-wave symmetry of the order 
parameter and a mass-enhancement factor 
$\lambda=0.02$.\cite{Friedl90} 
Not all Raman experiments could be consistently described within this 
picture leading to an intensive discussion at that 
time.\cite{McCarty92,Thomsen91,Altendorf92} 
More recent theoretical treatments of phonon renormalization have 
considered $d_{x^2-y^2}$ symmetry of the order 
parameter.\cite{Nicol93,Devereaux94b,Normand96} 
The latter theories have confirmed that the matching symmetries of 
phonon and 
gap are important to obtain strong phonon renormalizations.
The largely varying $2\Delta/k_BT_c$ ratios obtained 
in the earlier phonon self-energy 
studies\cite{McCarty92,Thomsen91,Altendorf92} 
can be accounted for if one not only considers the symmetry but also 
the strong doping dependence of the superconducting gap.

Except for some recent investigations of mercury or carbonate based 
four-layer compounds,\cite{Hadjiev98b,Hadjiev98a} most of the self-energy 
studies have been carried out on the RE-123 system.
In contrast, $2\Delta$ peaks, which are a signature of pair-breaking 
excitations, have been investigated in various cuprate families such as 
the single-layer compounds $\rm La_{2-x}Sr_xCuO_{4-\delta}$ 
(Ref.\onlinecite{Chen94}) and
$\rm Tl_2Ba_2Cu_1O_{6+\delta}$ 
(Tl-2201),\cite{Zaitsev95,Nemetschek93}
or the double-layer compounds 
Bi-2212 (Ref.\onlinecite{Blumberg97,Kendziora95,Hackl98}) or 
Y-123 (Ref.\onlinecite{Chen93,Cooper90,Thomsen91b}).
Above $T_c$ a structureless background is observed
in all plane-polarized excitation 
geometries.\cite{Reznik93}
The background can be assigned to incoherent quasi-particle 
scattering.\cite{Varma89,Ruvalds92,Jaklic97}
This scattering, which is observed 
with other spectroscopic methods as well,\cite{Ruvalds-Review} 
extends at least up to energy transfers of 
$\sim 1$ eV.\cite{Reznik93}
Except for very high dopings, 
low-temperature Raman spectra of the aforementioned compounds share the 
following features: (i) The energy of the $\rm B_{1g}$ $2\Delta$ peak 
exceeds that of the $\rm A_{1g}$ $2\Delta$ peak and increases with 
decreasing doping. (ii) The intensity of the 
background features increase with increasing doping.

Recently, various calculations have been presented which allow a description 
of the low-temperature electronic continuum in different polarization 
geometries.\cite{Devereaux94,Devereaux95a,DevereauxAdmit,Krantz95,Jiang96,Strohm97}
These theories rely on the approach of Klein and
Dierker\cite{Klein84} in which the mass part of the interaction of light
with superconducting electrons of a conventional electron gas exhibiting
a gap in its excitation spectrum is used. 
The main features observed in the Raman spectra, such as 
different energies and power laws in different polarization 
geometries, can be described within a $d_{x^2-y^2}$-wave 
picture, at least near optimal doping.\cite{Devereaux94} 
However, the relatively strong intensities of the $\rm A_{1g}$ $2\Delta$ 
peaks,\cite{Chen93} 
the vanishing intensities of the gap features at low doping 
levels,\cite{Chen97,Nemetschek97,Hackl98} and 
the converging peak energies at high doping 
levels\cite{Kendziora95,Bock98}
are only three features that cannot be explained with the present theories. 
Hence, it appears that the strong electron-electron interactions 
have to be more thoroughly considered in theoretical descriptions 
of the Raman spectra.\cite{Manske97,Manske98}

Whereas previous works focussed either on the gap features or on the 
phonon self-energy effects, here we investigate how far both effects are 
correlated.
In order to do that we use an extended Fano formula that we have 
presented previously\cite{Bock99} that takes into 
account the real and imaginary parts of the electronic response 
function and a phenomenological description of the response.
This work extends our previous study
towards lower dopings and includes investigations of the 
Raman responses in $\rm A_{1g}$ and $\rm B_{2g}$ 
symmetry in addition to the ones in $\rm B_{1g}$.
The phenomenology that we have presented allows us to 
obtain measures of the 
electron-phonon coupling, e.g. via the mass-enhancement 
factor $\lambda$.
Moreover, it enables us to identify the shape of the electronic 
background in the presence of strongly interacting phonons
which is of particular interest.

The investigated epitaxial $\rm Y_{1-x}(Pr,Ca)_xBa_2Cu_3O_{6+y}$ 
films cover a wide doping range.
To access the underdoped side of the phase diagram
we use oxygen reduction as well as Pr substitution.
Comparison of both methods allows us to study the influence of broken 
chains and of additional scatterers close to the CuO$_2$ planes on 
the Raman response.
We also compare the results of ordered samples with others in which 
site-substitution disorder with respect to 
a Pr $\leftrightarrow$ Ba or a Ca $\leftrightarrow$ Ba 
exchange is present.
We complete our study with an investigation of a single crystal.

In addition to the Raman spectra we have studied the transport 
properties of the films which
facilitates a determination of $T_c$ and $T^*$.
The three main interests of this study are:
(i) We want to investigate the doping dependence of the $2\Delta$ 
peaks in $\rm A_{1g}$ as well as in $\rm B_{1g}$ symmetry
in view of the non-mean field behavior of the gap observed in Bi-2212 
compounds, paying 
attention to the influence of disorder on these quantities.
(ii) We want to explore the temperature and doping dependence of 
the electron-phonon coupling and the phonon self-energies at 
high dopings, when $2\Delta$ approaches the phonon energy, and at low 
dopings, in the pseudogap regime, when $2\Delta$ peaks in $\rm 
B_{1g}$ are not observed.
At low dopings we are particularly interested to find signatures of 
the pseudogap.
(iii) We want to establish the temperature dependence of the gap 
energy and investigate whether we find indications for a monotonically 
closing pseudogap above $T_c$ as observed in ARPES measurements.
In our study
we benefit from the high temperature accuracies that are 
achieved when thin films are investigated.

This paper is organized as follows. 
In Sec. \ref{sec:exp} properties of the investigated films and the 
single crystal are presented as well as the setups that have been used 
for the Raman and transport measurements.
After a brief recapitulation of the phenomenology used to describe 
phonons superimposed on an interfering background we will 
give an example how the data are analyzed in Sec. \ref{sec:anal}.
In Sec. \ref{subsec:films} the influence of the oxygen content on the 
electronic background as well as on the phonon self-energy effects
in thin films is discussed. 
Results obtained on a single crystal are given in Sec. \ref{subsec:crystals}. 
The influence of the element substitutions Pr and Ca for Y on the 
background feature as well as on the phonon self-energy effects is 
presented in Sec. \ref{sec:subst}.
Whereas in Sec. \ref{subsec:order} ordered samples are investigated, 
Sec. \ref{subsec:disorder} focusses on the effects of 
site-substitution disorder when the Ba site is substituted in 
addition to the rare-earth site.
In Sec. \ref{subsec:dop} analyses of Raman and transport data are used 
to assign particular doping levels to the films and the single crystal.
Also the pseudogap temperatures will be determined.
Subsequently, in Secs. \ref{subsec:pse} and \ref{subsec:elba}, 
we will discuss our results 
regarding the phonon self-energy effects and the 
background features.
Peculiarities of the electronic responses are given in 
Sec. \ref{subsec:pec}.
A discussion of electron-phonon coupling and intensity anomalies 
of the $\rm B_{1g}$ phonon is found in Sec. \ref{subsec:epc}.
Conclusions are drawn in Sec. \ref{sec:con}.

\section{experimental details}
\label{sec:exp}

\subsection{Samples}
\label{subsec:sam}
We study $\rm Y_{1-x}(Pr,Ca)_xBa_2Cu_3O_{6+y}$ films 
with different doping levels achieved by substituting Y 
partially with Pr or Ca or by 
changing the chain-oxygen content y.
Pr substitution and oxygen reduction allows us to reduce the doping 
while Ca substitution allows us to obtain higher dopings.
{\em We want to emphasize at this point that
our Pr-123 films are reproducibly grown insulating antiferromagnets 
with well-defined two-magnon excitations.}\cite{Dieckmann96}
All films are grown by pulsed laser deposition 
on SrTiO$_3$(100) substrates and important properties 
are listed in Table \ref{tabProp}.
While the substituted films are grown in a standard process, which is 
optimized to obtain high oxygen contents 
$\mathrm{y}\approx 1$,\cite{Schilling93,Heinsohn98}
we used reduced oxygen-partial pressures during cool down for 
the oxygen-reduced ones.
Whereas the growth of the $\rm Y_{1-x}Pr_xBa_2Cu_3O_{6+y}$ films 
has been optimized to obtain homogeneous and smooth films with 
low precipitate densities,\cite{Dieckmann96} the Ca-doped films
exhibit somewhat poorer surface qualities, and a local variation of 
the Ca content from the nominal values cannot be excluded.
However, also for these films
the temperatures at which the superconductivity-induced features 
vanish are in good agreement with the transition temperatures.
In the Pr- and the Ca-doped films a 
site-substitution disorder with respect to the Ba site 
cannot completely be ruled out.
This disorder is promoted by the similar radii of the Ca, the Pr, and 
the Ba ions.
Even in laser-deposited Y-123 films it has been shown that 
$\rm Y \leftrightarrow Ba$ site-substitution can occur.
This disorder manifests itself 
via an increased $c$-axis parameter.\cite{Ye94}
Consequently, the phonon frequencies soften which can be used as an 
indication for the presence of disorder.
Except for two films, which will be analyzed in Sec. 
\ref{subsec:disorder}, all others are highly ordered 
exhibiting well-comparable properties and systematic doping 
dependencies. 

All films in this study have high epitaxial qualities with 
large $c$-axis oriented 
fractions $\delta_c \geq 92$ \% and high degrees of in-plane 
orientation $Q_c \geq 92$ \% determined by means of Raman 
spectroscopy as described elsewhere.\cite{Dieckmann96,Dieckmann95}
The high quality is a prerequisite to study 
plane-polarized excitation geometries.
In order to avoid surface contamination and oxygen depletion the 
films have been investigated shortly after preparation.

Given in Fig. \ref{figRes} are the temperature dependencies of the 
resistivities of the films.
Except for the film \#Ca2 we observe decreasing resistivities
with increasing doping as expected.
The resistivities are slightly higher than those of single crystals 
[for Y-123 single crystals $\rho(100K)=50$ $\mu\Omega$cm is a typical 
value\cite{Ito93,Takenaka94}].
The increased values may indicating additional sources of 
quasi-particle scattering.
However, the temperature dependencies, i.e. the slopes of the 
resistivity curves, are remarkably similar to single crystal 
data.\cite{Ito93,Takenaka94}
At high temperatures all films exhibit linearly decreasing 
resistivities. 
Deviations appear at lower temperatures, especially, in the 
underdoped films. 
This will be used to determine the pseudogap temperatures of the 
films in Sec. \ref{subsec:dop}.
The transition temperatures, defined by zero resistance, range 
from $T_c=56$ K in the underdoped film \#Ox1 up to $T_c=89.8$ K 
in the almost optimally doped film \#Ox3 and back down to 
$T_c=82.7$ K in the overdoped film \#Ca2.
The transition widths lie between $\Delta T_c\leq 10$ K at low and 
 $\Delta T_c\leq 2$ K at high dopings.

The $\rm YBa_2Cu_3O_{6+y}$
single crystal that has been investigated in this study has been 
grown in a zirconia crucible by a flux method described 
elsewhere.\cite{Liang92}
Prior to its investigation the surface has been etched in 1 \% 
bromine in methanol in order to remove possible contaminants.
Subsequently, the crystal was heated for several
hours under 1 bar oxygen-partial pressure 
in order to ensure a maximum oxygen content close to y=1.
According to its oxygen content a $T_c$ of $90\pm 1$ K is expected
for this crystal.\cite{Altendorf93}

\subsection{Setup}
\label{subsec:set}
For the Raman and transport investigations of the thin films they were 
mounted on the cold finger of a closed-cycle cryostat.
The electrical measurements are performed in van-der Pauw 
geometry using lock-in techniques in the same setup.
The crystal was mounted on the cold finger of a liquid He cryostat.
For excitation the Ar$^+$ laser lines at 458 nm and 514 nm were used 
in the studies of the thin films and the single crystal, respectively.
The thin film and single crystal studies have been performed in 
different setups described elsewhere,\cite{Ruebhausen97,Kaell94}
with the spectral resolutions (HWHM) set to 3 cm$^{-1}$ 
and 2 cm$^{-1}$, respectively.
{\em In the thin film study usage of the laser line, which was always recorded 
together with the Raman spectra, ensures a high relative frequency 
accuracy of 
$\approx 0.2$ cm$^{-1}$.}
All spectra have been corrected for the spectral response of 
spectrometer and detector
Spectra of the films \#Ox1 and \#Ox4 are additionally corrected 
in order to remove substrate signals.
They appeared due the poorer ratio between the thickness
(see Table \ref{tabProp}) and the penetration depth
(Ref. \onlinecite{Kircher91}) in these films.
In the thin film study we have used a power density of 110 Wcm$^{-2}$ 
and an effective spot radius of 40 $\mu$m leading to a typical heating 
of 2 K in the laser spot at 100 K.\cite{Bock95} 
The crystal has been investigated with power densities of 
120 Wcm$^{-2}$ and 270 Wcm$^{-2}$ 
and an effective spot radius of 20 $\mu$m leading to typical 
heatings of 4 K and 9 K in the laser spot at 100 K, respectively.\cite{Bock95} 
All given temperatures are spot temperatures.
$\rm B_{1g}$ and $\rm B_{2g}$ spectra are taken in polarization 
geometries which are denoted $z(x'y')\overline{z}$ and 
$z(xy)\overline{z}$ in Porto notation. 
Primed letters point out that the direction of polarization 
and the [100] axes of the sample form an angle of 45$^\circ$. 
$\rm A_{1g}$ spectra are obtained by subtracting the $\rm B_{2g}$ 
data from those measured in $z(x'x')\overline{z}$ geometry which give 
the $\rm A_{1g}$+$\rm B_{2g}$ responses.
In the thin film study the $x(a)$ and $y(b)$ axes cannot be 
distinguished due to twinning we therefore used the coordinate system 
given by the crystal axes of the substrate to describe the polarization 
geometries.
In case of the single crystal no $z(xy)\overline{z}$ spectra have been taken.

\section{Analysis of Raman spectra}
\label{sec:anal}

\subsection{Model}
\label{subsec:mod}

Apart from phonons, Raman spectra of cuprate superconductors 
consist of electronic (intraband) and magnetic (two-magnon) excitations. 
The latter dominate 
in the antiferromagnetic regime while the former gain spectral weight 
at higher dopings.\cite{Ruebhausen97}
When a phonon, characterized by its bare phonon frequency $\omega_p$ 
and its bare phonon linewidth $\Gamma$ (HWHM),
couples to intraband excitations, described by an 
electronic response function 
$\chi^e(\omega)=R^e(\omega)+i\varrho^e(\omega)$, 
its line shape becomes asymmetric and its self-energy is renormalized.
In case of the $\rm B_{1g}$ phonon in Re-123 compounds simultaneous 
descriptions which consider the interacting background have been 
obtained in a microscopic approach by Devereaux {\em et al.}
\cite{Devereaux95Pho} and a
phenomenological one by Chen {\em et al.}\cite{Chen93}
We have shown recently that both approaches resemble each other 
to a large extend and have obtained the following expression
for the imaginary part of the Raman response 
function ${\rm Im}\chi_{\sigma}(\omega)$ which is accessible in Raman 
experiments in $\sigma$ symmetry:
\begin{eqnarray}
\mathrm{Im}\chi_{\sigma}(\omega)=\gamma_{\sigma}^2
\varrho^e_{\sigma}(\omega)+
\frac{\gamma_{\sigma}^2 g_{\sigma}^2}{\gamma(\omega)
\left[1+\epsilon^2(\omega)\right]} \: \times 
\label{eqRamChi}
\end{eqnarray}
\begin{displaymath}
  \left\{ [R^e_{\sigma}(\omega) + R_{pp}]^2 - 
  2\epsilon(\omega)[R^e_{\sigma}(\omega) + R_{pp}]\,
  \varrho^e_{\sigma}(\omega) 
  -\varrho^e_{\sigma}(\omega)^2 \right\} \, ,
\end{displaymath}
with the renormalized phonon parameters 
$\gamma(\omega)=\Gamma+g_{\sigma}^2\varrho^e_{\sigma}(\omega)$ and 
$\epsilon(\omega)=\left[\omega^2-\omega^2_{\nu}(\omega)\right]
/[2\omega_p\gamma(\omega)]$ with 
$\omega_{\nu}^2(\omega)=\omega_p^2-2\omega_p 
g_{\sigma}^2R^e_{\sigma}(\omega)$.
In the above expression
$\gamma_{\sigma}$ represents the symmetry elements of the 
electron-photon vertex projected out by the incoming and outgoing 
polarization vectors, 
$g_{\sigma}$ is the lowest order expansion coefficient of the 
electron-phonon vertex describing the coupling to non-resonant intraband 
electronic excitations which is independent of momentum but 
may vary with temperature,
and $R_{pp}=g_{pp}/(\gamma_{\sigma}\cdot g_{\sigma})$ with
$g_{pp}$ a constant representing 
an abbreviated ``photon-phonon'' vertex which describes
the coupling to resonant interband electronic 
excitations.\cite{Devereaux95Pho}
{\em In the following we will refer to $g_{\sigma}$ as the effective 
electron-phonon coupling constant.}
The measured Raman intensity $I_{\sigma}(\omega)$ is related to 
the imaginary 
part of the electronic response function via the fluctuation-dissipation 
theorem:\cite{Hayes78}
\begin{eqnarray}
  I_{\sigma}(\omega)=\left[1+n(\omega)\right]I_{0,\sigma}(\omega)
  = A\left[1+n(\omega)\right]{\rm Im}\,\chi_{\sigma}(\omega)\, ,
  \label{eqFDT}
\end{eqnarray}
with $n(\omega)$ the Bose factor, $I_{0,\sigma}(\omega)$ 
the Raman efficiency, 
and $A$ a proportionality constant.
In order to arrive at a formula that allows us to model the Raman 
spectra we make the assumption that
{\em the phonon is only renormalized by 
the Raman-active electronic response}
$\gamma_{\sigma}^2\chi^e_{\sigma}(\omega)$, neglecting all 
other self-energy contributions 
like, e.g. those due to anharmonic phonon-phonon interactions
which are therefore included in $\omega_p$ and $\Gamma$.
This assumption allows us to express the Raman 
efficiency in the following way:\cite{Bock99}
\begin{eqnarray}
    I_{0,\sigma}(\omega)=\varrho_*(\omega)+\frac{C}{\gamma(\omega)
	\left[1+\epsilon^2(\omega)\right]} 
\label{eqRamInt}
\end{eqnarray}
\begin{displaymath}
 \times \left\{ \left[\frac{R_{tot}(\omega)}C\right]^2 - 
  2\epsilon(\omega)\frac{R_{tot}(\omega)}C\frac{\varrho_*(\omega)}C - 
  \left[\frac{\varrho_*(\omega)}C\right]^2 \right\} \, , 
\end{displaymath}
with the constant $C=A\gamma_{\sigma}^2/g_{\sigma}^2$ and the 
substitutions 
$\varrho_*(\omega)=Cg_{\sigma}^2\varrho^e_{\sigma}(\omega)$ and 
$R_{tot}(\omega)=R_*(\omega)+ R_0$ with 
$R_*(\omega)=Cg_{\sigma}^2R^e_{\sigma}(\omega)$ 
and $R_0=Cg_{\sigma}^2R_{pp}$.

Equation (\ref{eqRamInt}) can be simplified to resemble the 
conventional Fano 
formula\cite{Cooper90}
\begin{eqnarray}
	I_{Fano}(\omega)=C_{Fano}
	\frac{[q+\epsilon_{Fano}(\omega)]^2}
	{[1+\epsilon^2_{Fano}(\omega)]}  
\label{eqFano}
\end{eqnarray}
by setting 
$\epsilon_{Fano}(\omega) =(\omega^2- \omega^2_{\nu})/
(\omega_{\nu}\gamma_p)
\simeq(\omega- \omega_{\nu})/\gamma_p$,
the asymmetry parameter 
$q=-R_{tot}(\omega_{\nu})/\varrho_*(\omega_{\nu})$, 
and the intensity $C_{Fano}=\varrho_*^2(\omega_{\nu})/(C \gamma_p)$, 
where $\omega_{\nu}=\omega_{\nu}(\omega_p)$ and 
$\gamma_p=\gamma(\omega_p)$. 
In reference to the conventional Fano mechanism\cite{Hadjiev98b} 
we find for the total and the bare phonon intensity 
$I_{tot}=\frac{\pi}{C} R_{tot}^2(\omega_p)$ and 
$I_{phon}=\frac{\pi}{C} R_0^2=\pi A g_{pp}^2$.
With the above description of a Fano-type line shape
measures of the electron-phonon coupling 
can be obtained in two independent ways. 
First, in relevant units, via the mass-enhancement factor $\lambda$ 
defined by 
$\lambda\cdot\omega_p = 2R_*(\omega_p)/C=
2g_{\sigma}^2R_{\sigma}^e(\omega_p)$.
And second, in arbitrary units, via the reciprocal fit parameter
$1/C=g_{\sigma}^2/(A\gamma_{\sigma}^2)$.
{\em For brevity we will drop the suffix $\sigma$ in the 
discussion of the electron-phonon coupling in the
following as we are only dealing with the Fano-type line shape of the 
$B_{1g}$ phonon in $B_{1g}$ symmetry in this work.}

So far, a phonon is described by the four parameters $\omega_p$, 
$\Gamma$, $R_0$, and $C$. 
In addition, the frequency dependent real and imaginary parts of the 
electronic response have to be known.
An expression that successfully models the 
imaginary and the real parts of electronic 
response function observed in cuprate superconductors is:\cite{Bock99}
\begin{equation}
  \varrho_*(\omega)=
  I_{\infty}\tanh\left(\omega/\omega_T\right) + 
  \label{eqRho}
\end{equation}
\begin{displaymath}
  \left[\frac{C_{2\Delta}}{1+\epsilon_{2\Delta}^2(\omega)} - 
  \left(\omega \rightarrow-\omega \right) \right] -
  \left[\frac{C_{sup}}{1+\epsilon_{sup}^2(\omega)} - 
  \left(\omega \rightarrow-\omega \right) \right] \: ,
\end{displaymath}
where $I_{\infty}$, $C_{2\Delta}$, and $C_{sup}$ are fit parameters for the 
respective intensities,
$\omega_T$ and $\omega_{2\Delta}$ are fit parameters for the
crossover frequency of the background and the position of the $2\Delta$ 
peak, and $\Gamma_{2\Delta}$ is the width (HWHM) of the $2\Delta$ 
peak.
For the Lorentzians the abbreviations
$\epsilon_{2\Delta}(\omega)=
(\omega-\omega_{2\Delta})/\Gamma_{2\Delta}$ and
$\epsilon_{sup}(\omega)=
(\omega-\omega_{sup})/\Gamma_{sup}$ are used. 
While the first models the $2\Delta$ peak the second describes the 
suppression of spectral weight observed for $\omega \rightarrow 0$.
The second terms in the brackets are necessary to fulfill the 
symmetry requirements for the Raman response. 
The Lorentzian modelling the suppression 
is bound to the $2\Delta$ peak by setting
$2\omega_{sup}=2\Gamma_{sup}=\omega_{2\Delta}-\Gamma_{2\Delta}$ in 
order to reduce the number of free parameters.
The hyperbolic tangent which is used to reproduce the flat 
incoherent response is arbitrarily cut off at 
$\omega_{cut}=8000$ cm$^{-1}$.
As a consequence, the Hilbert transformed incoherent part contains a 
constant error.\cite{KKRvonTanH}
Due to this error the bare phonon frequency $\omega_p$ 
and the resonant intensity contribution $R_0$ cannot be 
determined accurately and depend on the chosen value of $\omega_{cut}$.
In detail, it turns out that 
$R_*(\omega_p)/C$ \{$R_0/C$\} increases \{decreases\} by
$I_{\infty}/C \cdot \frac{2}{\pi} 
\mathrm{ln}$~$(\omega_{cut}^{new}/\omega_{cut}^{old})$ when the 
cut-off frequency is varied.
With a typical value of $I_{\infty}/C=1$ at optimal doping 
corrections of $\sim 0.2$ cm$^{-1}$ appear when $\omega_{cut}$ is 
increased from 8000 cm$^{-1}$ to 11000 cm$^{-1}$.
Whereas the absolute values of the bare phonon frequency and the bare 
phonon intensity have to be considered with some care for the reason 
given above, the relative changes at different temperatures are not 
affected.

Equations (\ref{eqRamInt}) and (\ref{eqRho}) allow us to describe a phonon 
together with the interacting background that renormalizes it. 
If several phonons are observed, each acquires an interference term as 
the one given in Eq. (\ref{eqRamInt}).
An illustration of the fit parameters is shown in Fig. \ref{figVis} 
where an artificial spectrum with a phonon superimposed is displayed.
In the left panel the efficiency $I_0(\omega)$, the interacting, and the 
non-interacting phonon contributions are shown.
In the right panel the imaginary part of the electronic response 
function $\varrho_*(\omega)$ that is used in the description of the 
interacting phonon contribution and the separate background fractions are 
depicted.
Instead of the fit parameter $\omega_{2\Delta}$ we will use the center 
frequency of the peak in the imaginary part of the electronic response 
function as a measure of the $2\Delta$ peaks in 
the respective symmetries.
{\em For simplicity we will refer to the imaginary part of the electronic 
response function in the following as the electronic response.}

\subsection{Example}
\label{subsec:exa}

To give an example we show in 
Fig. \ref{figAllGeom} the efficiencies $I_{0}(\omega)$ of the 
film \#Ox4 in $\rm A_{1g}$, $\rm B_{1g}$, and $\rm B_{2g}$ symmetry, 
measured well above and below the transition.
We have normalized the 
$\rm B_{1g}$ efficiency of the film \#Ox4 at 18 K to unity 
above 650 cm$^{-1}$. 
{\em After consideration of the laser power and the scan time,
we use the same scaling factor for the spectra of all other films, 
geometries, and temperatures.
No further manipulations are needed to obtain well-comparable spectra.}
In the order of their energy we observe in Fig. \ref{figAllGeom}
the Ba mode at 120 cm$^{-1}$, the Cu(2) mode at 150 cm$^{-1}$,
the $\rm B_{1g}$ phonon which is
the out-of phase plane oxygen mode O(2)--O(3) at 340 cm$^{-1}$, 
the in phase plane oxygen mode O(2)+O(3) at 430 cm$^{-1}$, 
and the
apical oxygen mode O(4) at 500 cm$^{-1}$. These are the 5 Raman-active 
$\rm A_g$ phonons of Y-123.\cite{Thomsen91Rev} 
Please note, that the appearance of the Ba and the O(4) mode in 
$\rm B_{1g}$ symmetry is not related to the presence of $a$-axis 
oriented grains, as these modes also appear in single-crystal data 
with equal strength.\cite{Devereaux98}
Only in perfect tetragonal samples the phonons have exclusively 
$\rm A_{1g}$ or $\rm B_{1g}$ character. In the orthorhombic phase, 
i.e. at higher dopings, no exact selection rules apply and $\rm A_{1g}$ 
modes can appear in $\rm B_{1g}$ symmetry and vice versa.
In addition to the allowed modes we always observe 
a weak feature around 590 cm$^{-1}$. This feature grows in the 
oxygen-reduced samples in which it is accompanied by other modes
at 230 cm$^{-1}$ and 280 cm$^{-1}$. 
The modes at 230, 280, and 590 cm$^{-1}$ are modes of 
Y-123-O$\rm_{6+y}$ dominated by vibrations of Cu(1) and 
O(1).\cite{Wake91,Thomsen92,Ivanov95} 
They become Raman-active for $0<\mathrm{y}<1$
due to disorder in the occupancy of the chain-oxygen site. 
As the 590 cm$^{-1}$ mode is also observed in fully oxygenated samples 
it is likely to assume that it also indicates other types of disorder, 
probably at the Ba site.

Whereas the Raman efficiencies in $\rm B_{2g}$ symmetry are not altered 
by the superconducting transition in the film \#Ox4,
as seen in Fig. \ref{figAllGeom}, 
two main features are observed in the other symmetries. 
Namely, an 
increasing background in $\rm A_{1g}$ symmetry around 300 cm$^{-1}$ 
and an increasing $\rm B_{1g}$ phonon intensity accompanied by a 
reduction of spectral weight below 250 cm$^{-1}$ in $\rm B_{1g}$ symmetry.
In order to identify the background features and to separate the 
phononic contribution the following description is used:
(i) We describe the background with the expression given in 
Eq. (\ref{eqRho}).
(ii) We describe strong Fano profiles, such as the $\rm B_{1g}$ 
phonon in $\rm B_{1g}$ (symmetry), and the Ba and the O(4) 
mode in $\rm A_{1g}$,
with the interference term of Eq. (\ref{eqRamInt}).
(iii) We describe weak Fano profiles, such as the Ba mode in $\rm 
B_{1g}$, and the $\rm B_{1g}$ phonon in $\rm A_{1g}$, with
Eq. (\ref{eqFano}).
(iv) We describe symmetric phonons, such as the 
Cu(2) mode in $\rm A_{1g}$, with Lorentzians.

The $\rm A_{1g}$ efficiency of the film \#Ox4 at 18 K and its description 
are given in the uppermost spectrum of Fig. \ref{figFit}(a).
Below, the modeled phononic signal is displayed, i.e. the 
Lorentzians and the simple and improved Fano profiles. 
Above $\sim 120$ cm$^{-1}$ the phononic signal becomes negative as a 
consequence of the destructive interference of Ba mode and background.
Subtracting the phononic signal from the efficiency we obtain the electronic 
background $\varrho_*(\omega)$ which is shown in the second spectrum 
from the bottom.
In the background 
a $2\Delta$ peak at $\sim 280$ cm$^{-1}$ as well as a monotonically 
decreasing intensity for $\omega \rightarrow 0$
can be identified.
At the bottom of Fig. \ref{figFit}(a) the real part of the electronic 
background 
$R_{*}(\omega)$ is shown.
In order to obtain $R_{*}(\omega)$ we have performed a numerical Hilbert 
transformation of $\varrho_*(\omega)$.
For the transformation the measured spectrum is taken as constant 
for high frequencies up to $\omega_{cut}$
and is interpolated to zero intensity at $\omega=0$; for negative 
frequencies the antisymmetry of $\varrho_*(\omega)$ has been 
used.
Evidently, the description of $R_{*}(\omega)$ used in the fit agrees well
with the numerically obtained data.

Similarly to Fig. \ref{figFit}(a) we show in Fig. \ref{figFit}(b) the 
results of the analysis of the $\rm B_{1g}$ efficiency of the film 
\#Ox4 at 18 K.
Due to the strong coupling of the $\rm B_{1g}$ phonon to the 
electronic background, as evidenced by the high asymmetry of the 
profile, there is a substantial ``negative intensity'' observable in 
the phononic signal right at the position of the $2\Delta$ peak.
This explains why the $2\Delta$ peak in $\rm B_{1g}$ symmetry is 
hardly seen in the raw data shown in Fig. \ref{figAllGeom}. 
In fact, a simple division of spectra in order to obtain a 
measure of the $2\Delta$ peak yields misleading results when strong 
Fano profiles are present.
The features seen in the electronic background in $\rm B_{1g}$ 
symmetry are a $2\Delta$ peak at $\sim 390$ cm$^{-1}$, 
i.e. higher than in $\rm A_{1g}$ symmetry, and a decreasing 
background at small Raman shifts which falls off with 
$\sim \omega^{2.4}$ from the peak down to 270 cm$^{-1}$
and with $\sim \omega^{1.0}$ at lower frequencies.

In Fig. \ref{figBackTemp} we present the temperature evolution of the 
backgrounds in $\rm A_{1g}$ and $\rm B_{1g}$ symmetry which have been 
obtained after subtraction of the phononic signals from the 
efficiencies as explained above. 
One can still identify the positions of the strongest phonons via 
the increased noise levels which grow as $\sqrt{I_0(\omega)}$.
In addition, a typical renormalized frequency of the $\rm B_{1g}$ 
phonon is indicated via a dashed-dotted line in the right panel. 
At the bottom the 18 K data are shown.
With increasing temperature we observe that the $2\Delta$ peaks 
broaden while they concomitantly shift to somewhat lower energies.
Above 57 K the background features have vanished in $\rm A_{1g}$ 
symmetry
while a weak bump around 300 cm$^{-1}$ remains up to 87 K in $\rm 
B_{1g}$ geometry.
In this geometry, the background feature passes
the $\rm B_{1g}$ phonon between 57 K and 77 K.
Above 87 K a completely flat background is observed above 200 cm$^{-1}$
in both geometries.
These backgrounds are almost entirely described by 
the incoherent contribution, expressed via the first term in 
Eq. (\ref{eqRho}), except for small broad humps at the 
positions where the $2\Delta$ peaks are observed close to $T_c$ which 
are hardly resolvable in Fig. \ref{figBackTemp}.

\section{oxygen doping}
\label{sec:ox-dop}
The simplest way to vary the doping in the cuprate superconductors is 
by changing their oxygen content.
Under useful conditions it is possible to vary only the oxygen 
content in the charge-reservoir layers where the binding energies are 
typically weaker than in the CuO$_2$ planes.
In principle, this can be done during the preparation choosing specific 
oxygen-partial pressures. 
Often the oxygen content is changed ex-situ, e.g. after single 
crystals have been separated from the melt. 
In the RE-123 system the charge reservoir is provided by the 
copper-oxygen chains.
During the laser deposition the 
$\rm Y_{1-x}(Pr,Ca)_xBa_2Cu_3O_{6+y}$ films 
are tetragonal with $\mathrm{y}\approx 0$. 
High oxygen contents of $\mathrm{y}\approx 1$ are achieved during cool down
in 1 bar oxygen.
In order to obtain the oxygen-reduced films this
pressure has been decreased to values which are given in Table 
\ref{tabProp}.

\subsection{Thin Films}
\label{subsec:films}
In Fig. \ref{figBackOx} we show the electronic responses 
$\varrho_*(\omega)$ of the oxygen-doped films 
in $\rm A_{1g}$ and $\rm B_{1g}$ 
symmetry obtained after subtracting the phononic 
excitations from the efficiencies.
As substantial substrate signal was present in the $z(x'x')\overline{z}$
spectra of the film \#Ox1 this background will not be discussed.
Similar to Fig. \ref{figBackTemp}, 
we find increased noise levels at the positions of the 
strongest phonons in Fig. \ref{figBackOx}.
In agreement with previous studies on single crystals of Y-123
(Ref. \onlinecite{Chen93}) and of Bi-2212 (Ref. \onlinecite{Kendziora95}) 
we find increasing energies and 
decreasing intensities of the $2\Delta$ peaks in $\rm B_{1g}$ 
symmetry with decreasing doping.
At the lowest doping in the film \#Ox1 ($\mathrm{y}\approx 0.52$) the $2\Delta$ peak 
has vanished. This was also observed in single crystals, 
where studies in $\rm A_{1g}$ symmetry revealed a vanishing
$2\Delta$ peak as well.
However, the $2\Delta$ peaks do not vanish equally fast in $\rm A_{1g}$ 
symmetry and comparable strengths are observed for the films \#Ox2,
\#Ox3, and \#Ox4. 
Another difference to the $\rm B_{1g}$ feature is that
the $\rm A_{1g}$ $2\Delta$ peak follows $T_c$ having a
a maximum value in the 
almost optimally doped film \#Ox3 with 
$\sim 305$ cm$^{-1}$.
Regarding the power laws we find different behaviors in 
$\rm A_{1g}$ and $\rm B_{1g}$ symmetry.
Whereas the $\rm A_{1g}$ response below $\sim 250$ is proportional to 
$\omega^{1.0}$ 
in all films, the $\rm B_{1g}$ responses for $\omega \rightarrow 0$ 
changes slope 
with exponents of $1.00$, $0.90$, $0.4$, and 
$0.2$ in the films \#Ox4, \#Ox3, \#Ox2, and \#Ox1, respectively. 
Whereas the gap features are strongly doping dependent in $\rm B_{1g}$ 
symmetry, the responses above 650 cm$^{-1}$
exhibit comparable intensities.
Similarly, we find that the $\rm A_{1g}$ responses are well-comparable
above 650 cm$^{-1}$.

The temperature dependencies of the fit parameters of the $\rm B_{1g}$ 
phonons are shown in Fig. \ref{figPhonOx} in which the doping 
increases from left to right.
Beside the bare phonon parameters and the self-energy contributions 
at $\omega=\omega_p$ also the renormalized frequencies 
$\omega_{\nu}$ and linewidths $\gamma_p$ 
are depicted which facilitates a comparison with previous data obtained 
with the simple Fano description according to Eq. (\ref{eqFano}).
Solid lines are fits to the anharmonic decay for both 
models,\cite{Hadjiev98b} and anomalies below $T_c$ are 
discussed with respect to these fits.
Looking at the renormalized phonon parameters we observe 
softenings in the films \#Ox2, \#Ox3, and \#Ox4 
in agreement with the relative 
positions of the $\rm B_{1g}$ $2\Delta$ peaks and the phonon energies.
No superconductivity-induced change is observed in the film \#Ox1 in 
which 
the $2\Delta$ peak is absent.
In the film \#Ox4, which exhibits a $\rm B_{1g}$ $2\Delta$ peak 
in the vicinity of the phonon we find a strong broadening of 5.3 cm$^{-1}$. 
In contrast, slight sharpenings of 0.5 cm$^{-1}$ and 0.8 cm$^{-1}$ are 
observed in the films \#Ox3 and \#Ox2, in which the $\rm B_{1g}$ $2\Delta$ 
peaks evolve at substantially higher energies.

Our fit procedure now allows us to identify the self-energy effects 
which originate from the background.
As we have already observed in a previous study,\cite{Bock99} we find
a convincing correlation
between the changes of the linewidth appearing below $T_c$ and 
the varying 
self-energy contributions $\varrho_*(\omega_p)/C$ for the 
$\rm B_{1g}$ phonon in RE-123 films.
It turns out that the slight sharpenings in the films \#Ox2 and \#Ox3 
are a consequence of reduced effective couplings at low 
temperatures, as the background intensities at the 
phonon position $\varrho_*(\omega_p)$ are hardly affected. 
The broadening in the film \#Ox4, however, is almost entirely 
described by a slightly increased intensity $\varrho_*(\omega_p)$ 
accompanied by a substantially increased effective coupling.

There are two different representations of the effective 
electron-phonon coupling in 
Fig. \ref{figPhonOx}. 
First, we show the real part of the self-energy 
$R_*(\omega_p)/C$ which is proportional to 
the mass-enhancement factor $\lambda$ with 
$\lambda=2R_*(\omega_p)/(C\cdot \omega_p)=
2g^2R^e(\omega_p)/\omega_p$.
Second, we display $I_{\infty}/C=g^2\varrho^e(\omega \rightarrow \infty)$ 
which is
the inverse fit parameter $C$ normalized by the background 
intensity measured at high Raman shifts $I_{\infty}$.
While $\lambda$ allows us to quantify the strength of the 
coupling in relevant units, $I_{\infty}/C$ is a more accurate measure, 
however, only in arbitrary units. 
It can be seen in Fig. \ref{figPhonOx} that both representations of 
the effective coupling closely follow each other and increase with 
increasing doping.
In detail, we obtain almost vanishing coupling in the film
\#Ox1 with $\lambda \rightarrow 0.004$ for low 
temperatures whereas $\lambda\approx 0.015$
in the films \#Ox2, \#Ox3, and \#Ox4 above $T_c$.
As already mentioned, somewhat decreasing couplings are observed in 
the films \#Ox2 and \#Ox3 below $T_c$.
In contrast, we find a remarkable strengthening of the coupling 
in the film \#Ox4 at low temperatures.

We now turn to the lowest panel in 
Fig. \ref{figPhonOx} in which
beside the total intensities $I_{tot}$ also the bare phonon 
intensities $I_{phon}$ are given.
Even though increasing intensities $I_{tot}$ are observed below $T_c$
in all films, they have different origins.
Whereas the increase in the film \#Ox4 is a result of an increasing 
background contribution $R_*(\omega_p)$ accompanied by an increasing 
effective coupling, the increase in the other films originate from 
increasing values of $R_0$ which compensate the decreasing effective 
couplings.
In detail, it turns out that the bare phonon intensity 
$I_{phon}$ shows a change of slope at low dopings and a minimum 
in the film \#Ox4 at temperatures which increase monotonically with 
decreasing doping.
It appears as if the effective coupling approaches a maximum value 
close to this temperature.
Moreover, we find that the onsets of the softening in $\omega_p$ 
and $\omega_{\nu}$ are correlated with this temperature 
which is e.g. more than 10 K above $T_c$ in the film \#Ox2.

\subsection{Single Crystals}
\label{subsec:crystals}
A similar study of the doping dependence of the 
background features and of the phonon self-energy effects has been 
carried out by Irwin and co-workers who have investigated Y-123 
single crystals with different oxygen contents 
grown by Liang {\em et al.}\cite{Liang92}
While Altendorf {\em et al.} focussed on the $\rm B_{1g}$ phonon over 
a wide doping range down to $\mathrm{y}\approx 0.7$ using the simple Fano 
formula given in Eq. (\ref{eqFano}),\cite{Altendorf93}
Chen {\em et al.} explored the $2\Delta$ peaks 
for $0.93\leq \mathrm{y} \leq 1$ using an 
expression similar to Eq. (\ref{eqRamInt}).\cite{Chen93}
In the work of Chen {\em et al.} a Green's function model based on the 
approaches of Nitzan\cite{Nitzan74} and Klein\cite{Klein75} is used in order 
to obtain a simultaneous description of the phononic and electronic 
excitations.
Being mainly interested in the electronic background,
Chen {\em et al.} considered an electronic response function with a 
frequency dependent imaginary and a constant real part.
Especially, they assumed that the real part is small.
Comparison with Fig. \ref{figFit} shows, that $R_{tot}(\omega)$ will only 
be constant for a sufficient energy range when 
$\omega_{\nu}>(2\Delta+\Gamma_{2\Delta})$ or 
$\omega_{\nu}<(2\Delta-\Gamma_{2\Delta})$.
Also, we find that $R_{tot}(\omega)$ is of the order or even exceeds
$\varrho_*(\omega)$ in contrast to the above assumption.
The backgrounds obtained in their study must therefore not be 
identical to the ones that would have been obtained using 
Eqs. (\ref{eqRamInt}) and (\ref{eqRho}).

In order to facilitate a comparison of our results with the previous 
works but also in order to identify possible differences between the 
behaviors of single 
crystals and thin films we have investigated a single crystal 
also grown by Liang {\em et al.}\cite{Liang92}
Spectra and backgrounds of the single crystal 
are shown in Fig. \ref{figSpecCrys}.
Here, efficiencies containing $\rm A_{1g}$+$\rm B_{2g}$ symmetry are 
shown, as we did not take measurements in $z(xy)\overline{z}$ geometry.
However, as the $\rm B_{2g}$ response is small comparison to the one in $\rm 
A_{1g}$ symmetry and as the superconductivity-induced changes are 
weaker (see Fig. \ref{figAllGeom}),
the spectra are dominated by the $\rm A_{1g}$ response which 
facilitates a comparison with the thin film data.
In order to make this comparison easier, 
we normalize the intensities such, that the $\rm B_{1g}$ 
efficiency above 700 cm$^{-1}$ approaches unity in the 
18 K spectrum and use the same factor for the other temperatures and 
the other geometry as well.
With this scaling factor it turns out that 
the intensity of the $\rm B_{1g}$ phonon is comparable to 
the intensities seen in the films at least in the normal state
(see Fig. \ref{figPhonOx}).
However, due to the better spectral resolution and also because the 
natural linewidth in long-range ordered single crystals is smaller than in 
the films, the $\rm B_{1g}$ phonon peak height to background ratio is 
stronger in Fig. \ref{figSpecCrys}(a) than the one seen in Fig. 
\ref{figAllGeom}.
As the single crystals have large untwinned domains, enhanced 
orthorhombic distortions are present and the Cu(2) and the 
O(2)+O(3) modes appear in $\rm B_{1g}$ efficiency
in addition to the Ba and the O(4) mode.
For the same reason a relatively strong $\rm B_{1g}$ phonon is 
observed in $\rm A_{1g}$+$\rm B_{2g}$ symmetry.
Similar observations have been made by Chen {\em et al.}\cite{Chen93}

Regarding the electronic backgrounds of the single crystal in 
$\rm A_{1g}$+$\rm B_{2g}$ and 
$\rm B_{1g}$ symmetry, which are shown in Fig. \ref{figSpecCrys}(b) we 
find similar features than in the films: 
A strong $2\Delta$ peak in 
$\rm A_{1g}$+$\rm B_{2g}$ symmetry with a linear slope 
of $\sim \omega^{1.1}$ at small Raman shifts and
a weaker $2\Delta$ peak in $B_{1g}$ symmetry which falls down 
from the peak with a power of $\sim \omega^{3.0}$ and has a reduced slope 
of $\sim \omega^{1.0}$ below 270 cm$^{-1}$. 
The intensities of the $2\Delta$ peaks in the crystal are 
distinctly stronger than those of the films.
Also, the energies of the $2\Delta$ peaks differ from those in the 
films with a $2\Delta$ peak of $\sim 335$ cm$^{-1}$ 
in $\rm A_{1g}$+$\rm B_{2g}$ symmetry and a 
$\rm B_{1g}$ $2\Delta$ peak of $\sim 435$ cm$^{-1}$.
Both values are higher than those observed in the Y-123 film \#Ox4 
indicating a lower doping level.
Again one can see that the strong anti-resonance inhibits 
a clear identification of the $\rm B_{1g}$ $2\Delta$ peak in the raw data 
of Fig. \ref{figSpecCrys}(a).

The self-energy effects of the $\rm B_{1g}$ phonon of the single 
crystal
are given together with those of the oxygen-doped films in 
Fig. \ref{figPhonOx}.
Due to the better signal to noise ratio and the better spectral 
resolution we used smaller symbols to indicate the 
smaller error bars.
The absolute values of the frequencies and linewidths differ somewhat 
from those observed in the films.
Smaller linewidth appear as a consequence of the higher spectral 
resolution used in this study.
In addition, they reflect the increased grain size, i.e. the long-range order, 
of the single crystal.
Increased frequencies are most likely related to the smaller 
$c$-axis parameter of fully oxygenated single 
crystals\cite{Altendorf93} compared to 
those of fully oxygenated thin films.\cite{Matijasevic91}
Even though the absolute frequencies and linewidths do not agree, the 
self-energy effects, the effective coupling, 
and the intensities are well-comparable.

Looking at the phonon-self energy effects depicted in Fig. \ref{figPhonOx}
it turns out that the bare phonon 
follows the expected anharmonic decay to a large extend. 
Only a small increase of the bare phonon linewidth exists close to
$T_c$, whereas a significant softening of $\omega_p$ appears at low 
temperatures.
While the strengths of the coupling in the crystal agrees with that of 
the film \#Ox4 at very high and low temperatures, the coupling in the 
crystal starts to increase already below 250 K exceeding the film 
value at $T_c$ by almost a factor of two with $\lambda\approx 0.023$.
The better signal to noise ratio allows us to identify clearly 
an intermediate 
maximum of 
the coupling appearing slightly below $T_c$ very clearly.
This maximum appears right at that temperature at which a small intermediate 
broadening in the renormalized phonon linewidth is observed and the 
bare phonon starts to soften.
Moreover, a distinct minimum of the bare phonon intensity $I_{phon}$, 
similar to the one seen in the film \#Ox4, appears in the same 
temperature regime.
The temperature at which the anomalies occur is higher than in the 
film \#Ox4 
which is another indication of a lower doping in the single 
crystal in view of the doping dependence of this 
temperature in the oxygen-doped films.
At low temperatures the total intensity is significantly 
stronger than those of the films. This is related to the strong intensity 
of the $\rm B_{1g}$ $2\Delta$ peak in the single crystal.

\section{element substitutions}
\label{sec:subst}
Many cuprate superconductors offer the possibility to vary the 
doping via the exchange of ions with different valences. 
Well-known exchange pairs are $\rm Sr^{2+}\leftrightarrow La^{3+}$, 
e.g. in the La-214 compounds, or $\rm Y^{3+}\leftrightarrow Ca^{2+}$, 
e.g. in the Bi-2212 compounds. 
The first allows one to dope holes while the second reduces the hole 
concentration.
Often this approach is limited as some materials 
become unstable when the dopant concentration exceeds a certain value. 
Stable Ca-substituted Y-123 compounds are obtained up to a dopant 
concentration of 0.5.\cite{Buckley91}
However, in these samples Ca also substituted the Ba site, and the 
relative fraction of Ca on the Y site dropped with increasing Ca 
content. For $\mathrm{x}\leq0.2$ less than 20 \% of the Ca substituted the Ba 
site.
$\rm Y_{1-x}Pr_x$-123 is a stable compound for all Pr concentrations 
up to pure Pr-123.
The reason why Pr substitution is varying the doping when replacing a 
rare-earth atom in a RE-123 compound is generally attributed to its 
capability to localize holes in hybridized orbitals of Pr $4f$ 
and adjacent O $2p$.\cite{Fehrenbacher93}
However, recently it has been suggested that substantial 
site-substitution disorder with respect to a 
$\rm Pr \leftrightarrow Ba$ exchange is inhibiting superconductivity 
in Pr-123.\cite{PRL-Pr-SC} 
In fact, the latter authors claimed to find superconductivity in 
samples in which Pr is exclusively occupying the rare-earth site.
Recently, however, it was argued that the small magnetic moments 
observed in the superconducting Pr-123 samples rather indicates that 
Pr is present at the Ba-site in a substantial volume 
fraction.\cite{PRL-Comm}
The high resistivity of the superconducting Pr-123 samples as well as 
the impossibility to prepare them with a high purity in large volume 
fractions, as was stated by the authors of 
Ref. \onlinecite{PRL-Pr-SC}, are further 
indications for the presence of strong site-substitution disorder in 
superconducting Pr-123.

As already stated above, our Pr-123 films are reproducibly grown 
insulating antiferromagnets.
However, in these films as well as in the Ca-substituted ones a 
site-substitution disorder appearing at the Ba and the RE site 
cannot completely be ruled out,
as the growth conditions during the laser deposition are away from 
equilibrium.
The appearance of disorder can depend on the morphology of the substrate 
surface or on slight changes in the preparation conditions which 
cannot be controlled in all details.
It manifests itself in decreased phonon frequencies and a weakened 
and broadened Ba phonon\cite{Bock99Ba} as noted in Sec. \ref{subsec:sam}.
In the following sections we will compare the Raman spectra of ordered 
and disordered films beginning with the former ones.

\subsection{Ordered Samples}
\label{subsec:order}
Figure \ref{figBackOrd} shows the electronic responses
$\varrho_*(\omega)$ of the ordered films \#Pr2,
\#Pr3, and \#Ca2 at 18 K in $\rm A_{1g}$ and $\rm B_{1g}$ symmetry.
For comparison with the oxygen-doped films, the energies of the 
$2\Delta$ peaks of the film \#Ox4 are given as dashed lines.
In case of the film \#Pr3 we did not measure the $\rm B_{2g}$ spectra 
directly.
However, as will be shown in Sec. \ref{subsec:elba}, the measured 
$\rm B_{2g}$ responses of the under- or slightly overdoped films varied hardly.
Therefore, we subtracted the $\rm B_{2g}$ spectra of the film \#Pr2 
from the $z(x'x')\overline{z}$ data of the film \#Pr3 in order to obtain its 
$\rm A_{1g}$ response.
The strengths of the background features in these films are 
comparable to those in the oxygen-reduced films.
Regarding the $\rm B_{1g}$ responses in Fig. \ref{figBackOrd} 
we observe a similar behavior as 
in the oxygen-doped films: With decreasing doping the $2\Delta$ peak 
shifts to increasing frequencies from $\sim 305$ cm$^{-1}$ in the 
film \#Ca2 up to $\sim 580$ cm$^{-1}$ in the film \#Pr3.
Concomitantly, we observe a change of slope at 
Raman shifts below 270 cm$^{-1}$ from $\sim \omega^{1.7}$ in the 
film \#Ca2 to $\sim \omega^{0.4}$ 
in the film \#Pr2.
Note, that while the $T_c$'s of the films \#Ca2 and \#Pr3 vary by 
only 4 \%, their $2\Delta$ peaks vary by almost a factor of two. 
Even though the $\rm B_{1g}$ responses of the similarly low-doped 
samples \#Pr2 and \#Ox1 are flat, 
showing no indication of a $2\Delta$ peak, 
the Pr-doped sample has a 50 \% stronger intensity at high Raman shifts.
Another remarkable difference to the oxygen-doped samples appears, 
if we look at the $\rm A_{1g}$ responses.
Here, we observe a strong $2\Delta$ peak even in the low-doped
sample \#Pr2. 
At similar dopings the $\rm A_{1g}$ $2\Delta$ peaks in oxygen-reduced 
samples have been observed to vanish.\cite{Chen97,Nemetschek97,Hackl98}
The energies of the $\rm A_{1g}$ $2\Delta$ peak follow again the 
$T_c$'s of the samples showing a maximum value of $\sim 325$ 
cm$^{-1}$ in the film \#Pr3.
Also, we find that the slopes at small Raman shifts are almost linear
with $\sim \omega^{1.05\pm0.15}$ similar 
as in the oxygen-doped films. 
The $\rm A_{1g}$ efficiencies at high Raman shifts, however, 
are significantly increased in the Pr-substituted films 
compared to the oxygen-doped ones.
These efficiencies even exceed the values observed in $\rm B_{1g}$ 
symmetry.

The temperature dependencies of the self-energy effects of all 
substituted films are depicted in Fig. \ref{figPhonSubst}
where the doping level increases from left to right.
Again the bare as well as the renormalized phonon parameters are given
together with the self-energy contributions at 
$\omega=\omega_p$.
Comparing the renormalized frequency of the ordered films 
with the expected values 
according to the anharmonic decays we observe 
softenings of 5 cm$^{-1}$ and 1.5 cm$^{-1}$ in the films \#Pr2 and 
\#Pr3, respectively, as well as a hardening of 1 cm$^{-1}$ in the 
film \#Ca2.
The frequency shifts are in agreement with the relative 
positions of the $\rm B_{1g}$ $2\Delta$ peaks and the phonon energies.
When comparing the renormalized linewidths of the ordered films 
with the anharmonic decays we observe a broadening of
5.4 cm$^{-1}$ in the film \#Ca2, in which the $\rm B_{1g}$ $2\Delta$ peak 
evolves in the vicinity of the phonon.
The broadening is well-comparable to the one observed in the film \#Ox4.
In the other films \#Pr2 and \#Pr3 we observe sharpenings of 
2.5 cm$^{-1}$ and 1.5 cm$^{-1}$, respectively, which are clearly 
stronger than the softenings in the oxygen-reduced films.
This holds especially for the film \#Pr2, as we did not observe any 
linewidth anomaly in the film with the lowest oxygen content \#Ox1.

We now relate the self-energy effects of 
the $\rm B_{1g}$ phonon in the ordered substituted films to the shape of and 
the coupling to the observed electronic responses.
It turns out that neither the bare frequencies $\omega_p$ nor the 
bare linewidth $\Gamma$ follow the anharmonic decay in any of the 
ordered films.
For instance, we find a sharpening of $\Delta\Gamma=-2$ cm$^{-1}$ in 
the film \#Pr2 and a broadening of $\Delta\Gamma=2$ cm$^{-1}$ in 
the film \#Ca2 as well as concomitant softenings and hardenings of 
$\Delta\omega_p=-2.5$ cm$^{-1}$ and $\Delta\omega_p=4$ cm$^{-1}$, 
respectively. 
As only weak or even vanishing $\rm B_{1g}$ $2\Delta$ peaks are 
observed in the Pr-doped samples (see Fig. \ref{figBackOrd}) 
it is not surprising that the strong 
softenings and sharpenings which appear below $T_c$ in these films 
cannot be attributed to the redistribution of the background.
However, in the film \#Ca2 a strong $\rm B_{1g}$ $2\Delta$ peak was 
observed and the remaining discrepancies are unexpected.

The two representations of the coupling 
$R_*(\omega_p)/C$ and $I_{\infty}/C$ given in 
Fig. \ref{figPhonSubst} are
similarly good correlated as those in Fig. \ref{figPhonOx}.
Also, we find that the effective couplings, which are well-comparable 
to the values observed in the oxygen-doped films, 
weaken with decreasing doping.
In the highest doped film \#Ca2 we observe an increasing effective 
coupling $I_{\infty}/C$ below $T_c$ similar as in the film \#Ox4, 
whereas decreasing 
effective couplings are observed in the other two films at low 
temperatures.
While increasing bare phonon intensities $I_{phon}$ are observed in 
the lowest row in the two Pr-doped films, a monotonic decrease 
is observed in the film \#Ca2 below $T_c$.
The latter is in agreement with a previous study of us,\cite{Bock98}
in which an increase of the total phonon intensity 
$I_{tot}$ was absent in films whose $2\Delta$ peaks are below the phonon 
frequency for all temperatures.
Similar as in the oxygen-doped films we find that the bare phonon intensity 
$I_{phon}$ shows a change of slope at low dopings and a minimum 
in the film \#Ca2 at those temperatures where a strong effective 
coupling is observed.
Moreover, an increase of this temperature with decreasing doping is 
observed in Fig. \ref{figPhonSubst} as well.

\subsection{Disordered Samples}
\label{subsec:disorder}
In order to study the influence of disorder on the Raman response and 
on the phonon self-energy effects we have chosen two pairs 
of films, one pair with a Pr content of 20 \% and one pair with a Ca content 
of 5 \%. 
The ordered films, which have been presented before, 
are \#Pr2 and \#Ca2, the disordered films are \#Pr1 and \#Ca1.
The $\rm A_{1g}$ response of the film \#Pr1 was obtained by 
subtracting the $\rm B_{2g}$ spectra of the film \#Pr2 
from its $z(x'x')\overline{z}$ data similar as for the film \#Pr3.
In Fig. \ref{figBackDis} the backgrounds of the two pairs of films are shown. 
Clearly, the disordered films exhibit weaker background features in 
$\rm A_{1g}$ as well as in $\rm B_{1g}$ symmetry.
In addition, we observe shifts of the $2\Delta$ peaks. 
It turns out, that the $\rm A_{1g}$ and $\rm B_{1g}$ $2\Delta$ peaks 
in the film \#Ca1 appear at $\sim 270$ cm$^{-1}$ and 
$\sim 350$ cm$^{-1}$, respectively, corresponding to increases of 
25 cm$^{-1}$ and 45 cm$^{-1}$ in comparison to the film \#Ca2.
The $\rm A_{1g}$ $2\Delta$ peak of the film \#Pr1, however, has a lower 
energy of $\sim 270$ cm$^{-1}$ compared to 
$\sim 290$ cm$^{-1}$ in the film \#Pr2.
Independent of the disorder we observe that $\rm A_{1g}$ and 
 $\rm B_{1g}$ responses merge above the $2\Delta$ peaks in the 
Ca-doped films whereas they are offset by some constant value 
in the Pr-doped ones. 

We now come to the self-energy effects of the disordered films which are 
depicted in Fig. \ref{figPhonSubst} in shaded columns.
Regarding the renormalized frequencies we find a softening in the 
film \#Ca1 whereas a hardening was observed in the film \#Ca2.
This correlates well with the position of the $\rm B_{1g}$ $2\Delta$ 
peak which exceeds the phonon frequency in the film \#Ca1 in contrast 
to the film \#Ca2.
Whereas a strong hardening of the bare phonon frequencies was observed 
in the film \#Ca2, it follows the anharmonic decay 
in the film \#Ca1.
In agreement with the decreased intensity of the background feature 
seen in Fig. \ref{figBackDis},
the broadening in the disordered film \#Ca1 is weaker than in the 
ordered one.
In a similar manner the sharpening in the film \#Pr1 is smaller than 
the one in the film \#Pr2.
The softenings of the bare and the renormalized phonon frequencies 
are more pronounced in the disordered film \#Pr1 than in the 
ordered.
Moreover, they appear at a remarkably high temperature of 150 K.
However, all other phonon frequencies soften below this temperature as 
well (not shown).
This indicates that the softening is rather a signature of a lattice 
instability occurring in this temperature range than a result of 
a particularly strong interaction with electronic excitations.

Comparing the strengths of the effective couplings between the 
disordered and the ordered film we find corresponding 
values with respect to both, $R_*(\omega_p)/C$ and $I_{\infty}/C$.
Whereas the maximum effective coupling in the film \#Ca1 is 
observed for $T\rightarrow 0$ similar as in the film 
\#Ca2, the maximum effective coupling in the Pr-doped 
disordered film is observed at somewhat higher temperatures near the 
temperature of the lattice instability.
Another difference between disordered and ordered films 
is related to the bare and total intensities of the $\rm B_{1g}$ phonon.
Clearly, $I_{phon}$ and $I_{tot}$ are smaller in the film \#Ca1 than 
in the film \#Ca2 down to 100 K. 
Below that temperature a superconductivity-induced 
increase is observed in $I_{tot}$ in the film \#Ca1.
This finding is in agreement with our observations in the ordered 
films, in which increasing intensities at low temperatures have only 
been observed when the energy of the $\rm B_{1g}$ $2\Delta$ peak 
for $T \rightarrow 0$ exceeds the phonon value.
An intensity anomaly is also observed in the film \#Pr1, however, the 
ratio $I_{tot}(18K)/I_{tot}(152K)$ is 35 \% smaller than in the 
film \#Pr2.

\section{discussion}
\label{sec:dis}

\subsection{Doping Levels and Pseudogap Temperatures}
\label{subsec:dop}
The doping level $p$ is defined as the number of holes per 
CuO$_2$ plane in the unit cell.
In order to determine $p$ for 
the $\rm Y_{1-x}(Pr,Ca)_xBa_2Cu_3O_{6+y}$ films 
two different methods are used.
The first method relies on a general $T_c(p)$ relation which was 
proven to be valid for various HTS compounds by 
Tallon {\em et al.}\cite{Tallon95}
\begin{equation}
 T_c(p) = T_{c,max}[1-82.6(p-0.16)^2] \, .
      \label{eqTc(p)}
\end{equation}
In our study we used a maximum transition temperature 
$T_{c,max}=90\pm 1$ K for the thin films taken from a recent 
systematic study.\cite{Heinsohn98}
The thus calculated doping values are given in Table \ref{tabProp}. 

The second method uses the Pr, Ca, or oxygen contents and specific 
$T_c(\mathrm{x})$ and $T_c(\mathrm{y})$ relations.
In case of the Ca-doped films the simple relation 
$p=0.187+\mathrm{x}/2$ has 
been given by Tallon {\em et al.}\cite{Tallon95} 
This relation relies on the 
fact that fully oxygenated Y-123 has a doping value of $p=0.187$ which 
was estimated in the same work.
For the Pr-doped films we use the above estimate for x=0 and take into 
account that $T_c$ vanishes for Pr contents $\mathrm{x}\simeq 0.5$
(Ref. \onlinecite{Radousky}) which corresponds to $p\simeq0.05$ 
according to Eq. (\ref{eqTc(p)}).
Consequently, we employ $p=0.187-0.274\mathrm{x}$ 
for the Pr-doped films.
The mean value of both methods $\bar{p}$
is used in the following
except for the disordered films \#Pr1 and \#Ca1.
For these two films we will only take into account 
the values calculated 
from the transition temperature as the disorder 
invalidates the other consideration.
Comparing the two disorder-order-pairs (\#Ca1 / \#Ca2 ) 
and ( \#Pr1 / \#Pr2 ) we find decreased doping levels in the 
disordered films.

To determine $p$ for the oxygen-doped films, the relation 
$p=-0.023+0.21\mathrm{y}$, valid for $0.45\leq \mathrm{y} \leq 1$, 
has been taken from Ref. \onlinecite{Tallon95}.
Application of this formula requires knowledge of the oxygen 
contents which have been determined as follows:
In case of the film \#Ox4 we can put y=1 as this film has been 
prepared with a high oxygen-partial pressure during the cool down.
For the other films we may use $T_c(\mathrm{y})$ relations obtained 
from published data.\cite{Ito93,Takenaka94}
With $T_c=56.0$ K we find $y=0.52\pm0.03$ for the film \#Ox1.
For the last two films an estimation of the oxygen content with the 
same method is not useful as their $T_c$'s are close to 
the optimal value of 90 K for our films.
In order to estimate their oxygen contents we made use of the 
intensity of the disorder-induced Raman mode at 230 cm$^{-1}$ which 
only appears when copper-oxygen chains are interrupted.\cite{Iliev97}
The relative intensities of this mode at 290 K are 100 \%, 51 \%, 
and 23 \% 
for the films \#Ox1, \#Ox2, and \#Ox3, respectively 
($\rm B_{1g}$ spectra at 18 K are shown in Fig. \ref{figBackB1g}).
In the presence of long-range order one would expect 
the strongest intensity of this mode near $\mathrm{y}\approx 0.75$, and
a decreased intensity for $\mathrm{y}\simeq 0.5$ where only
every second chain is filled in the ortho II phase.\cite{Jorgensen90}
In fact, a recent investigation of oxygen-reduced detwinned or 
slightly twinned Y-123 single crystals is in agreement with this 
expectation.\cite{Kaell98}
However, as a result of the lattice mismatch to the substrate the 
investigated films here are heavily twinned inhibiting a long-range 
order of the chain oxygen. 
Assuming therefore randomly occupied chain-oxygen sites\cite{Furrer94}
and considering only a central and two neighboring unit cells we have
calculated the probability of interrupted chains as a function of the 
chain-oxygen content y.\cite{Int-230(y)}
In this calculation a maximum probability for an interrupted chain is found
at $\mathrm{y}\approx 0.6$.
With the knowledge of the oxygen content of the film \#Ox1 we could
estimate the oxygen contents and, as a result, the doping values of the films 
\#Ox2 and \#Ox3 which are given in Table \ref{tabProp}. 
Similar as in the Pr- and Ca-doped films 
we will use the mean value $\bar{p}$ from the dopings 
determined via Eq. (\ref{eqTc(p)}) and via the oxygen contents 
in the following.

As has been noted in Sec. \ref{subsec:sam}, the resistivity curves 
given in Fig. \ref{figRes} deviate from the linear behavior
$\rho(T)= \rho(0) + \alpha T$ below a 
certain temperature.
Similar observations have been made in single crystal studies and the 
temperature at which the deviation occurs 
has been identified as the pseudogap temperature 
$T^*$.\cite{Ito93,Takenaka94}
In order to show the different $T^*$ of the films it is advisable to use 
the representation of the resistivity shown in Fig. \ref{figPseudo}.
Here, the resistivities are normalized by the slope $\alpha$ of the 
$T$-linear part after $\rho(0)$ has been subtracted.
We use a 1 \% criterion to determine $T^*$ from these graphs,
meaning that $[\rho(T^*)-\rho(0)] /(\alpha T^*) \approx X-0.01$.
Here, $X$ is the saturation value of the normalized resistivity at 
low temperatures which is in the regime $1\pm 0.005$ in this study.
The estimated pseudogap temperatures are given in Table \ref{tabProp}.
We find that the ordered films exhibit increasing pseudogap temperatures 
with decreasing doping.
The two disordered films \#Pr1 and \#Ca1 clearly differ from this 
behavior with $T^*=100\pm 5$ K and $T^*=145\pm 10$ K, respectively.
Additional scattering contributions or structural instabilities are two 
possibilities that could be responsible for the differences.
In case of the film \#Ox1 we find deviations from the linear 
$T$-dependence of the resistivity
almost in the entire temperature range studied.
As this hinders an estimation of $T^*$,
its value is given in brackets in Table \ref{tabProp}.

At comparable dopings oxygen-reduced films have significantly 
higher pseudogap temperatures than Pr-substituted ones.
This could be related to a stronger $c$-axis coupling present in the 
Pr-doped films in which the copper-oxygen chains remain unaltered.
It could also be related to the presence of the hole-localizing Pr 
atoms right between the planes which may act as strong scatters and 
thereby hinder the formation of the pseudogap.
In case of the oxygen-doped films a comparison with published single 
crystal data\cite{Ito93,Takenaka94} can be carried out.
It shows that the $T^*$ of the films are higher than those of the 
crystals which are below 220 K for $\mathrm{y}\geq 0.78$.
This could again be related to a decreased $c$-axis coupling as a 
greater disorder in the occupation of the chain-oxygen site can be 
expected in the strongly twinned films.
According to published results\cite{Ito93,Takenaka94} a pseudogap 
temperature of $\sim 370$ K would be expected at the oxygen content 
of the film \#Ox1 which 
is in agreement with our failure to estimate one in this film.

\subsection{Phonon Self-Energies}
\label{subsec:pse}
In the upper panel of Fig. \ref{figDopSE} we have depicted the differences 
between the renormalized linewidths for $T \rightarrow 0$ 
and the values expected from the anharmonic decays of the 
investigated films and the single crystal 
versus energy of their $\rm B_{1g}$ $2\Delta$ peaks.
In the lower panel the corresponding graph for the 
renormalized frequencies is given.
Beside the results of this study also single crystal data obtained by 
Altendorf {\em et al.}\cite{Altendorf93} are included.
The corresponding $\rm B_{1g}$ $2\Delta$ peaks are taken from the 
work of Chen {\em et al.}\cite{Chen93} who investigated the same
crystals.
Together with other results of our analyses,
the self-energy effects obtained in this study are also given in Table
\ref{tabAnal}. 

Regarding the upper panel of Fig. \ref{figDopSE} we find broadening 
or sharpening in samples in which the $2\Delta$ peaks are below or 
above $\sim 500$ cm$^{-1}$.
This indicates that the electronic response, or the coupling to it, or 
both, weaken in the superconducting state when the gap is above that 
value.
The broadening appears to be at maximum when phonon frequency and 
$2\Delta$ peak match each other.
This is expected, as the phonon will decay most effectively via 
breaking of a Cooper pair under this condition.
That the broadening does not vanish at higher dopings, i.e. at lower 
peak energies, is a consequence of the $d$-wave shape of the gap 
function.\cite{Nicol93,Normand96}
The disordered film \#Ca1, whose self-energy effects are shown in 
brackets, exhibits a smaller broadening than the two adjacent ordered 
films.
This is directly related to the weaker $\rm B_{1g}$ $2\Delta$ peak 
observed in the film \#Ca1.
Another sample that seems to deviate from the general trend is the 
fully oxygenated single crystal investigated by Altendorf and Chen.
It exhibits a strong broadening of 5.9 cm$^{-1}$ and a high peak 
energy of 460 cm$^{-1}$.
As the renormalized phonon parameters do not depend strongly on the 
applied description one may speculate that the energy of the $2\Delta$ 
peak is overestimated in the description employed by Chen 
{\em et al.} in Ref. \onlinecite{Chen93}.
This could be a consequence of the approximation made in their work
in which the frequency dependence 
of the real part of the electronic response function $R_*(\omega)$
has been neglected.
In fact, if we compare the phonon parameters of the fully oxygenated 
single crystal of their study with those of the one we studied we 
find at least two indications that their crystal might have a 
higher oxygen content.
Knowing that the $c$ axis shrinks with increasing oxygen 
content,\cite{Jorgensen90} and that the intermediate broadening, or 
equivalently, the maximum asymmetry of the $\rm B_{1g}$ phonon 
shifts to lower temperatures with increasing oxygen 
content\cite{Altendorf93} these indications are:
(i) That the renormalized $\rm B_{1g}$ phonon frequency at $\sim 100$ K is 
1 cm$^{-1}$ higher in their crystal indicating a slightly smaller 
$c$ axis.
(ii) That the temperature at which they observe the maximum intermediate 
broadening and the maximum asymmetry is $\sim 70$ K and therefore 10 K 
lower as in our crystal.
As a result of the higher oxygen content we would expect an increased 
doping and therefore a $\rm B_{1g}$ $2\Delta$ peak below 435 cm$^{-1}$
in their fully oxygenated single crystal compared to the one we studied.

Turning now to the lower panel of Fig. \ref{figDopSE} we find hardening 
or softening for all samples whose $2\Delta$ peaks are below or 
above the phonon frequency.
This corresponds well to the result of a simple picture when the 
interaction is treated in second order perturbation theory.
When the center frequency of the background peak is above the phonon 
it is shifted downwards and vice versa.
A symmetric background peak right on the phonon will yield a vanishing
frequency shift.
In Fig. \ref{figDopSE} the ordered as well as the disordered sample follow 
the same trend.
In the present work only the ordered film \#Ca2 shows hardening.
However, we have recently revealed the same behavior in disordered 
Ca-doped samples with higher Ca contents.\cite{Bock98}
In 10 \% and 15 \% doped films we observed $\rm B_{1g}$ $2\Delta$ peaks 
of 315 cm$^{-1}$ and 220 cm$^{-1}$ and accompanying hardenings of 
$\sim 0.2$ cm$^{-1}$ and $\sim 2$ cm$^{-1}$, respectively.

Whereas phonon self-energy effects are often given for a fixed 
electronic response and varying phonon frequencies,
the representation chosen in Fig. \ref{figDopSE} is of more practical 
relevance for at least two reasons.
First, because it is easier to vary the doping and therefore the 
$\rm B_{1g}$ $2\Delta$ peak by changing the oxygen content then to 
vary the phonon frequency, and second because 
the $2\Delta$ peak 
shifts over a wider energy range than the phonon does.
The observed self-energy effects of the $\rm B_{1g}$ phonon
principally resemble those seen in theoretical 
studies.\cite{Nicol93,Normand96}
For example we find a maximum in the imaginary part of the 
self-energy in that energy range where the real part has its maximal 
slope.
However, a more specific comparison with theoretical works is not possible
at the moment, as the self-energy effects are a consequence of
a coupling to the electronic 
response function for which a microscopic description is not available 
at present. 
Moreover, we have observed that the self energy-effects are not 
solely bound to the superconductivity-induced changes of the 
electronic excitation spectrum. 
Rather, they appear also in samples which do not exhibit $2\Delta$ 
peaks in the superconducting state and at temperatures above $T_c$ 
which will be discussed in the following section.
For example, we find a sharpening of 2.5 cm$^{-1}$ and a softening of
1.5 cm$^{-1}$ in the film \#Pr2 with an onset around 120 K, where 
$T_c$ is only 72 K.

\subsection{Electronic Background}
\label{subsec:elba}
{\em The high quality of our fit procedure shows up when 
$\rm A_{1g}$ and $\rm B_{1g}$ responses are plotted in the same graph 
as done in Figs. \ref{figSpecCrys} and \ref{figBackDis}.
Evidently, both responses are well-comparable above the $\rm B_{1g}$
$2\Delta$ peak in the single crystal and the Ca-doped films, 
even though the strongest phonons appear at 
substantially different energies in both symmetries. 
This unpredicted behavior is seen in all films whose dopings are
above the optimal value.
It gives us additional confidence in the obtained fit parameters.}

The doping dependence of the $2\Delta$ peaks in $\rm A_{1g}$ and
$\rm B_{1g}$ symmetry measured at 18 K is depicted 
in Fig. \ref{fig2DeltaDop}.
Beside our data we have again included those obtained by 
Chen {\em et al.}\cite{Chen93} in the graph.
In order to determine the doping of their crystals the oxygen contents 
as well as the $T_c$'s given in the their work have been used in a 
similar way as described in Sec. \ref{subsec:dop}.
Given as solid lines are the gap energies that would correspond to a 
mean field\cite{Won94}
approach with $2\Delta(p)=4.28$ $\mathrm{k}_B$$T_c(p)$ 
where $T_c(p)$ has the 
parabolic shape given in Eq. (\ref{eqTc(p)}).
Note the reduced energy scale in $\rm A_{1g}$ symmetry.

In agreement with a previous doping-dependent study of Bi-2212 single
crystals by Kendziora and Rosenberg\cite{Kendziora95}
we find that the $\rm A_{1g}$ 
$2\Delta$ peaks follow - more or less - a parabolic dependence 
being slightly above that particular curve.
However, with the aid of the Pr-doped films our study extends their results 
towards lower dopings.
Noteworthy, the energies of the $\rm A_{1g}$ $2\Delta$ peaks are almost 100 
cm$^{-1}$ lower in the RE-123 samples compared to those of the 
Bi-2212 single crystals. 
For example, Kendziora and Rosenberg observe $2\Delta$ peaks of 
$\sim 400$ cm$^{-1}$ near optimal doping.
The origin of this behavior is unknown at present.
The energies of the $\rm A_{1g}$ $2\Delta$ peaks of the single crystals 
shown in Fig. \ref{fig2DeltaDop} appear to 
be higher than those of the thin films.
However, both groups of samples exhibit a significant scatter of the 
data.
Please note also that the $\rm A_{1g}$ $2\Delta$ peaks of the single 
crystals given in Fig. \ref{fig2DeltaDop} 
are in fact $\rm A_{1g}$+$\rm B_{2g}$ $2\Delta$ peaks as the 
$\rm B_{2g}$ responses have not been subtracted.

Regarding the doping dependence of the $\rm B_{1g}$ $2\Delta$ peaks, 
we find again that the peak energies in the films are smaller than 
those of the crystals at comparable dopings.
The slope of the doping dependence of the film data, however, 
is well-comparable to 
the one observed in Bi-2212 single crystals using SIS tunneling 
and ARPES which is given as a dashed line.\cite{Miyakawa98}
This demonstrates a close relationship of the gap values in the Bi-2212 and 
the RE-123 system that has not been seen so far as high-quality tunneling 
or ARPES data cannot be obtained in the RE-123 system due to surface 
problems. 
Note that in this geometry data at low dopings are not available as 
the redistributions become unresolvable slightly below the optimal 
value. 

So far, we have not discussed the $\rm B_{2g}$ responses of the films.
This is related to our observation that they are not altered with 
temperature (see, e.g. Fig. \ref{figAllGeom}).
In addition, the $\rm B_{2g}$ responses are almost doping independent 
as shown in Fig. \ref{figBackB2g}(a).
In the determination of the the $\rm A_{1g}$ 
efficiencies in those two films \#Pr1 and \#Pr3 in which we did not 
measure the $\rm B_{2g}$ spectra we benefitted from this property.
Only in the Ca-doped film with the highest doping \#Ca2 we observe 
the evolution of a $\rm B_{2g}$ $2\Delta$ peak as 
shown in Fig. \ref{figBackB2g}(b).
The energy of $\sim 265$ cm$^{-1}$ is between those of 
the $\rm B_{1g}$ and $\rm A_{1g}$ $2\Delta$ peaks (see Table 
\ref{tabAnal}) in 
agreement with results obtained on similarly doped Bi-2212 single 
crystals.\cite{Kendziora95,Hackl98}
The intensities of the $\rm B_{2g}$ responses are distinctly smaller than 
those of the $\rm B_{1g}$ responses with ratios of 0.5 and 
0.3 at $\sim 700$ cm$^{-1}$
in the oxygen- or Pr-doped films and in the film \#Ca2, respectively.
Near optimal doping the observed ratios of $\sim 0.5$ 
are in agreement with those measured on Y-123 single 
crystals.\cite{Chen97,Krantz95}

We now turn to the doping dependence of the low-temperature electronic 
responses in $\rm A_{1g}$ and $\rm B_{1g}$ symmetry. 
Depicted in Fig. \ref{figBackDop} are the responses of two underdoped 
films \#Ox1 and \#Pr2, the almost optimally doped film \#Ox3, and the 
overdoped film \#Ca2 measured at 18 K.
The $\rm A_{1g}$ response of the film \#Ox1 is not given as it 
contained significant substrate signal as noted above.
In order to improve the signal/noise ratio we have 
subsumed 10 neighboring data points of the original curves 
yielding, however, a poorer spectral resolution.
Three main features can be read from Fig. \ref{figBackDop}:
(1) In contrast to oxygen reduction or Ca doping, Pr doping leads to 
an increased intensity in the entire energy range.
(2) The intensity of the $\rm A_{1g}$ $2\Delta$ peak is almost 
independent of the doping provided that filled chains are present.
(3) With decreasing doping the $\rm B_{1g}$ $2\Delta$ peak
rapidly looses its strength while shifting to higher energies.

(1) The incoherent background observed in Raman but also in 
other measurements, like e.g. optical conductivity or resistivity, 
can be assigned to inelastic scattering of quasi-particles in the 
CuO$_2$ plane.\cite{Varma89,Ruvalds92}
In this sense the enhanced intensities observed in Pr-substituted 
samples indicate an enhanced scattering rate.
In view of the specific electronic configuration of the Pr$^{3+}$ ion,
which can localize holes in the Fehrenbacher-Rice 
state,\cite{Fehrenbacher93} and in view 
of its position right in the middle of the copper-oxygen 
bilayer, an enhanced scattering probability may not be surprising. 
Interestingly, this enhancement is only observed in $\rm A_{1g}$ and 
$\rm B_{1g}$ symmetry and not in the $\rm B_{2g}$ 
spectra shown in Fig. \ref{figBackB2g}(a).
Whereas in $\rm B_{2g}$ symmetry the FS is predominantly sampled 
close to the diagonals $\pm k_x=\pm k_y$, the former two have 
significant contributions from the region close to 
$(\frac{\pi}{a},0)$.\cite{Devereaux94,Devereaux95a}
In view of this behavior our observation suggests that Pr doping 
may rather amplify already present inelastic scattering near the hot 
spots of the FS near $(\frac{\pi}{a},0)$
than add a new scattering channel.

(2) The observation of a rather strong $\rm A_{1g}$ $2\Delta$ peak in 
the underdoped film \#Pr2 is in contrast to the behavior seen in
$\rm B_{1g}$ symmetry.
More importantly, it is in contrast to the behavior of 
the $\rm A_{1g}$ responses of oxygen-reduced 
Y-123 single crystals\cite{Chen97,Nemetschek97,Hackl98} or under- up 
to almost optimally doped 
Bi-2212 single crystals\cite{Quilty98,Ruebhausen98Bi}
where $\rm A_{1g}$ $2\Delta$ peaks have not been observed.

There is a long ongoing debate on the understanding of the gap 
features in $\rm A_{1g}$ symmetry observed in cuprate superconductors.
Whereas Devereaux {\em et al.} initially claimed to obtain a 
simultaneous description of the electronic responses in $\rm A_{1g}$,
$\rm B_{1g}$, and $\rm B_{2g}$ symmetry in a one-band model using 
Fermi surface harmonics to expand the mass part of the interaction 
of light with matter for calculations 
along the Fermi line,\cite{Devereaux94,Devereaux95a}
they had to admit later that their results depend strongly 
on the expansion coefficients of the electron-photon 
vertex which cannot be obtained unambiguously.\cite{DevereauxAdmit}
The same observation was made in a recent theoretical approach in 
which an integration over the entire Brillouin zone was carried 
out.\cite{Manske97,Manske98}

Early on Cardona and co-workers pointed out that the 
remarkable high intensity of the $\rm A_{1g}$ response indicates that 
multiple sheets of the FS are involved in the scattering 
process.\cite{Krantz95,Strohm97}
With multiple bands unscreened mass fluctuations can be expected in 
$\rm A_{1g}$ symmetry whereas strong screening is expected 
in a one-band model.
In this picture
it is difficult to understand why the gap features have 
similar shapes in different compounds even though the band structures 
vary significantly.
Another argument against the multiple-band picture was the 
observation of rather strong $\rm A_{1g}$ $2\Delta$ peaks in the 
single layer Tl-2201 close to 
optimal doping.\cite{Gasparov99,Zaitsev95,Gasparov98}
However, at higher dopings almost vanishing intensities of the $\rm A_{1g}$ 
feature and a persisting strong $\rm B_{1g}$ $2\Delta$ peak 
have been observed in Tl-2201.\cite{Gasparov99,Zaitsev95}
This is in agreement with calculations in which the electronic 
properties are dominated by a single band. 

In our study we still have filled chains in the Pr-substituted 
underdoped films. 
Hence, we keep the chain bands which can still provide 
additional sheets of the FS.
This could explain the unaltered intensities of the 
$\rm A_{1g}$ $2\Delta$ peaks
seen in the spectra of the Pr-doped samples. 
That high quality single crystals of optimally doped Bi-2212 
($T_c\approx 95$ K) do not show an $\rm A_{1g}$ $2\Delta$ 
peak\cite{Ruebhausen98Bi} in 
contrast to high quality single crystals of optimally doped Y-123 
($T_c\approx 93$ K)\cite{Chen93} is 
another support for our finding that chain-related bands may play a 
significant 
role for the Raman response in $\rm A_{1g}$ symmetry.
If we consider the $\rm A_{1g}$ response to be influenced by 
the presence of chains, it is likely to assume that the type of 
oxygen-ordering present in the chain layer will be of some relevance for 
the shape of the response.
The scatter in the energies of the $2\Delta$ peak close to optimal
doping seen in Fig. \ref{fig2DeltaDop} could be related to this 
particular order.
Moreover, one might interpret the significantly higher intensities of 
the $2\Delta$ peaks seen in the study of the
single crystal (see Fig. \ref{figSpecCrys}) as evidence for an 
increased oxygen-order in the chain layer in view of the 
disorder-related reduction of $2\Delta$ peak intensities shown in 
Fig. \ref{figBackDis}.

(3) When the intensities of the $2\Delta$ peaks decrease 
with decreasing doping, a structureless 
background remains in $\rm B_{1g}$ symmetry.
In contrast to the film \#Pr2, where additional background intensity 
is observed as already discussed, the oxygen-reduced film \#Ox1 
exhibits comparable intensities at low ($\sim 100$ cm$^{-1}$) and 
high ($\sim 800$ cm$^{-1}$) Raman shifts, showing a slightly rising 
background.

The nature of the low-energy Raman response has been the subject of 
various experimental and theoretical investigations.
Concerning $\rm B_{1g}$ symmetry the following observations have been 
made: 
(i) Whereas the low-energy $\rm B_{1g}$ response in the antiferromagnetic 
precursors can be attributed to two-magnon 
excitations in a good approximation,\cite{Ruebhausen97,Knoll90}
the two-magnon excitations loose and incoherent excitations gain 
strength with increasing doping.\cite{Dieckmann96,Ruebhausen97}
(ii) In optimally or slightly underdoped Bi-2212 single crystals the 
slope for $\omega \rightarrow 0$ cannot be described in a 
Drude-like picture in contrast to the response in $\rm B_{2g}$ 
symmetry.\cite{Einzel96}
(iii) In slightly overdoped Y-123-O$\rm_{6+y}$ 
single crystals with $\mathrm{y}\approx 1$
the temperature dependence of the $\rm B_{1g}$ 
response at small Raman shifts is not simply following 
the behavior expected in the frameworks of the 
marginal or the nested 
Fermi liquids.\cite{Reznik95}
(iv) The response, which does not depend on the excitation energy in 
moderately doped superconductors, becomes resonantly enhanced at 
very high doping levels exhibiting a resonance position similar to that 
observed for two-magnon excitations at lower 
doping levels.\cite{Kang96,Ruebhausen99}

In view of the above results it has been suggested\cite{Ruebhausen97} 
that the $\rm B_{1g}$ response consists at least of three different 
contributions in the investigated doping range. 
A magnetic contribution due to overdamped two-magnon excitations, which 
persist in the superconducting state, 
an electronic contribution resulting from incoherent quasi-particle 
scattering, and a mixed spin-charge response which appears below the 
pseudogap temperature. 
Recently, we have observed that magnetic scattering vanishes at high 
dopings $p\geq 0.21$ in Bi-2212 single crystals.\cite{Ruebhausen99}
Here, we didn't investigate samples with such high dopings, therefore
we can assume that two-magnon excitations are always present in the 
$\rm B_{1g}$ efficiencies at low temperatures, 
having increasing weight at lower dopings. 
Apparently, when the magnetic contributions strengthen below 
optimal doping the $\rm B_{1g}$ $2\Delta$ peaks vanish concomitantly 
in RE-123 compounds.
In Bi-2212 compounds weak pair-breaking-like excitations have been 
observed 
even at low dopings.\cite{Blumberg97,Kendziora95,Quilty98}
They appear around 600 cm$^{-1}$ and it was reported that they are 
also present above $T_c$.\cite{Blumberg97,Quilty98}
A detailed investigation of that frequency region is hindered 
in this study by the presence of a disorder-induced infrared-active 
phonon at 590 cm$^{-1}$ (see Fig. \ref{figBackB1g}).
In contrast to a previous study of a Y-123-O$_{6.5}$ single 
crystal,\cite{Chen97} the $\rm B_{1g}$ responses at large 
Raman shifts above $700$ cm$^{-1}$
remain almost doping independent in the oxygen-doped films
down to y=0.52 in the film \#Ox1 in our study.
This may be related to the fact that we had to subtract substrate 
signal in this film,
as we cannot exclude residual substrate intensity.
In fact, two strong substrate modes are still observed in the 18 K 
efficiency (see Fig. \ref{figBackB1g}).

\subsection{Peculiarities of the Electronic Response}
\label{subsec:pec}
In this section we will consider two peculiarities of the 
electronic response unveiled in this study.
One of these is the doping dependence of the exponent of the 
low-frequency power laws 
in $\rm A_{1g}$ and $\rm B_{1g}$ symmetry and another the influence of 
disorder on the strength and energy of the $2\Delta$ peaks.

(i) As shown in Table \ref{tabAnal} the $\rm A_{1g}$ responses fall of 
as $\omega^{1.0\pm 0.15}$ and no systematic dependence on the doping 
is observed.
In contrast, the exponents observed in $\rm B_{1g}$ symmetry 
increase systematically with doping.
In detail, we observed a slope $\sim \omega^{0.2}$ in the 
underdoped film \#Ox1, $\sim \omega^{0.9}$ in the 
almost optimally doped film \#Ox3, and $\sim \omega^{1.7}$ in the 
overdoped film \#Ca2.
In slightly overdoped films even a change of slope was observed 
appearing around 270 cm$^{-1}$ (see Secs. \ref{subsec:exa} 
and \ref{subsec:crystals}).
Above that frequency higher exponents of $\omega^{3.0}$ and 
$\omega^{2.3}$ are observed in the film \#Ox4 and the single crystal 
\#sc, respectively.
A similar crossover has been observed in a previous investigation of 
the electronic responses in differently doped Y-123 single 
crystals.\cite{Chen93}

The relevant power laws in Raman spectra of cuprate 
superconductors are those at small Raman shifts.
They may be compared with theoretical estimates obtained in 
$\omega/(2\Delta)$ expansions of the calculated electronic responses.
Assuming  $d$-wave symmetry of the order parameter
Devereaux {\em et al.}\cite{Devereaux94,Devereaux95a}
obtained power laws $\sim \omega$ and 
$\sim \omega^{3}$ in $\rm A_{1g}$ and $\rm B_{1g}$ symmetry, 
respectively.
Those have in fact been observed in case of optimally doped Bi-2212 
and identified as support for the $d$-wave 
scenario.\cite{Devereaux94,Devereaux95a}
The power laws of the $\rm B_{1g}$ responses in our study
clearly deviate from the value expected for a $d$-wave superconductor.
The deviation and its doping dependence cannot be accounted for with 
present theories and pose some constraints on any models of the Raman 
response.

(ii) The influence of disorder on the strength and 
energy of the $2\Delta$ peaks has not been established before.
The reduced intensities observed in this study are in 
agreement with theoretical expectations.\cite{Devereaux95,Bille99}
However, a quantitative analysis of the strength of the  
scattering rates lies beyond the scope of this work.
Beside diminished intensities we also observed shifts of the energies 
of the $2\Delta$ peaks when comparing the disordered with ordered 
films (see Table \ref{tabAnal}).
Whereas the peaks shifts towards higher energies in the slightly 
overdoped film \#Ca1, 
a down shift of the $\rm A_{1g}$ $2\Delta$ peak is observed in the 
underdoped 
film \#Pr1 in which a $\rm B_{1g}$ $2\Delta$ peak could not be identified.
Based on the doping dependence of the $2\Delta$ peaks shown in Fig. 
\ref{fig2DeltaDop} one can assign decreased doping levels to the
disordered films compared to the ordered ones.
This assignment agrees with their respective 
critical temperatures which have already been interpreted as 
indications for reduced doping levels in Sec. \ref{subsec:dop}.

\subsection{Electron-Phonon Coupling and Intensity Anomalies}
\label{subsec:epc}
Our description of phonon and interacting background according to 
Eqs. (\ref{eqRamInt}) and (\ref{eqRho}) allows us to obtain the 
mass-enhancement factor $\lambda$ as well as a measure of the 
effective electron-phonon coupling constant $g$.
They are determined from the fit parameters via the relations 
$\lambda=2R_*(\omega_p)/(C \cdot\omega_p)$ and 
$g^2\propto I_{\infty}/C$.
In contrast to $I_{\infty}/C$, the quantity $\lambda$ is not useful 
in the superconducting state as additional self-energy contributions 
due to the pair-breaking excitations appear. 
However, in the normal state both quantities are clearly correlated as seen in 
Figs. \ref{figPhonOx} and \ref{figPhonSubst} where instead of 
$\lambda$ the ratio
$R_*(\omega_p)/C$ is given.
Around optimal doping and at 100 K we find 
$R_*(\omega_p)/C=2.5$ cm$^{-1}$ 
yielding $\lambda=0.015$ when a typical bare phonon frequency 
$\omega_p=344$ cm$^{-1}$ is taken.
This value is slightly below $\lambda=0.021$ calculated 
by Rodriguez {\em et al.}\cite{Rodriguez90} using a
linear muffin-tin-orbital frozen-phonon calculation.
Somewhat higher values of the coupling with $\lambda=0.056$
are used in the work of 
Devereaux {\em et al.}\cite{Devereaux95Pho} in order to explain 
the line shape of the $\rm B_{1g}$
phonon in slightly overdoped Y-123.
In contrast to our approach, they use a microscopic description of the 
imaginary part of the
electronic response, taking impurity and electron-electron scattering 
into account as proposed in the theory of the nested Fermi 
liquid.\cite{Ruvalds92}
For their description of the background three fit parameters are required 
instead of the two, $I_{\infty}$ and $\omega_T$, we use.
Moreover, they describe the phonon using five fit parameters while 
we only have four: $\omega_p$, $\Gamma$, $C$, and $R_0$.
Their fit formula is identical to our
Eq. (\ref{eqRamInt}) in the limit of a strongly interacting phonon 
but differs in the case of an only weakly interacting, symmetric one.
An additional difference between both descriptions is that
Devereaux {\em et al.} do not consider the frequency dependent real 
part of the electronic response in their description but use $\lambda$ as an 
independent fitting parameter in order to describe the asymmetry 
of the phonon.
As their spectra are well-comparable to our thin 
film and single crystal data, the origin of the deviating coupling 
strengths is not obvious. 
It will most likely be related to the different degrees of freedom 
allowed in both approaches.

In our study a decreasing coupling is observed
in the oxygen-doped films with decreasing doping. 
In detail, we find $R_*(\omega_p)/C=2.54$ cm$^{-1}$ and 
$R_*(\omega_p)/C=0.09$ cm$^{-1}$ in the films \#Ox4 and \#Ox1 at 100 K,
corresponding to $\lambda=0.015$ and $\lambda=0.0005$, respectively.
The same trend has recently been observed in a single crystal study 
by Opel {\em et al.}\cite{Devereaux98} in which the microscopic approach of 
Devereaux {\em et al.}\cite{Devereaux95Pho} was employed.
However, they need again significantly higher values of the coupling to 
describe their spectra, even though these spectra
are well-comparable to our thin film data.
To give an example, they use $\lambda=0.026$ to describe the $\rm 
B_{1g}$ spectrum of an underdoped Y-123-O$\rm _{6.5}$ single 
crystal which exceeds our estimate by a factor of 50.
In addition to the higher degree of freedom allowed in the 
approach of Devereaux {\em et al.}\cite{Devereaux95Pho},
this difference may also be related to the somewhat different descriptions 
used in the limit of weakly interacting, symmetric phonons.
Clearly, this large discrepancy demands to be resolved.

In contrast to oxygen reduction, Pr substitution yields only weakly 
decreasing coupling constants as seen in Fig. \ref{figPhonSubst}.
However, at high Pr concentrations, which are not shown here, 
the $\rm B_{1g}$ phonon becomes symmetric, and 
$\lambda$ approaches zero in Pr-123 films.\cite{Dieckmann96}

To illustrate the doping dependence of the electron-phonon coupling in 
oxygen- and Ca-doped films in the
the superconducting state we show in 
Fig. \ref{figBackB1g} the Raman efficiencies in $\rm B_{1g}$ symmetry 
measured at 18 K.
The doping increases from top to bottom and the determined couplings 
$I_{\infty}/C$ are given for each spectrum.
Whereas we observe a strong, sharp, and symmetric $\rm B_{1g}$ phonon 
in the film \#Ox1, the mode is weak, broad, and asymmetric in the film 
\#Ca2.
Apparently, the disorder-induced modes at 230 cm$^{-1}$ and 590 cm$^{-1}$ 
appear and grow with decreasing oxygen content. 
This has been used in 
Sec. \ref{subsec:dop} to determine the oxygen contents of the films
\#Ox2 and \#Ox3, where, however, room-temperature data have been 
considered as these modes exhibit reduced intensities (bleaching) at 
lower temperatures.\cite{Wake91,Kaell98}

The decreasing effective couplings at low dopings as observed in
Figs. \ref{figPhonOx}, \ref{figPhonSubst}, or \ref{figBackB1g}, 
have to be discussed with respect to the assumptions 
which have been made in Sec. \ref{subsec:mod} in order 
to arrive at the description of the Raman efficiency in Eq. (\ref{eqRamInt}). 
There, we have explicitly assumed that the phonon gets renormalized by 
the electronic response function whose imaginary part is observed in the Raman 
spectrum.
As mentioned in Sec. \ref{subsec:elba} we can assume that the electronic 
response in $\rm B_{1g}$ symmetry consists of incoherent electronic 
and two-magnon excitations as well as a spin-charge response.
Especially, the two-magnon contribution will grow with decreasing doping.
The symmetric appearance of the $\rm B_{1g}$ phonon in the 
antiferromagnetic precursors Y-123-O$_6$ (Ref. \onlinecite{Burns91}) or 
Pr-123 (Ref. \onlinecite{Dieckmann96})
indicates that this mode is not interacting with two-magnon excitations, 
at least not with those at low energies which originate from the vicinity 
of the Brillouin zone.
The reduced effective coupling at low dopings 
therefore expresses at least partly the overestimation 
of the background contribution responsible for the observed 
self-energy effects.
Support for this picture is our observation that we find 
mass-enhancement factors which correspond fairly well to the results of 
the LDA-type calculation of Rodriguez {\em et al.}\cite{Rodriguez90}
at higher dopings, where the two-magnon excitations diminish.
However, in addition to the above considerations it is likely to 
assume that the density of states at the hot spots, which are sampled 
in $\rm B_{1g}$ symmetry, will affect the electron-phonon coupling as 
well. 
The density will increase strongly at high dopings as the van Hove 
singularity moves closer to the Fermi level.
Another origin for reduced effective coupling strength 

The fact, that the effective coupling of Pr-doped films is stronger at 
comparable dopings than the one in oxygen-reduced films may 
indicate that the low-energy magnetic contributions are growing weaker 
with Pr-substitution.
It may also be related to an enhancement of the coupling strength 
induced by the Fehrenbacher-Rice state of the Pr atom.

During the presentation of the self-energy effects we have stressed 
the finding that the effective coupling $I_{\infty}/C$ appears to be 
related to the bare phonon intensity $I_{phon}$ in the sense that 
both show anomalies at similar temperatures.
In detail, we find that $I_{\infty}/C$ is maximal near the temperature 
at which $I_{phon}$ is minimal at higher dopings or 
changes slope at lower dopings.
As stated earlier 
$I_{phon}=\frac{\pi}{C} R_0^2$ is proportional to $g_{pp}^2$, where $g_{pp}$
describes the coupling of the phonon to interband electronic 
excitations. 
$I_{\infty}/C$, on the other hand, is proportional to $g^2$, i.e. to the 
coupling of the phonon to intraband electronic excitations.
Our observations therefore indicate an intimate relation between both 
excitation channels.
To demonstrate this relation we display the ratio 
$(I_{\infty}/C)/I_{phon}=I_{\infty}/(\pi R_0^2)\propto (g/g_{pp})^2$ 
versus temperature in Fig. \ref{figTanomaly}.
This ratio reflects the asymmetry of the phonon line shape:
The higher the ratio, the higher the asymmetry.
For each sample the obtained ratios are normalized to their maximal 
values.
The fat vertical bar represent the temperatures at which the ratio is 
at maximum and to which we will refer as $T_{anomaly}$ in the 
following.
Neither in the highest doped film \#Ca2 nor in the one with the 
lowest doping \#Ox1 a clear maximum is observed.
In the former (latter) we find that the ratio grows (decreases) 
continuously with decreasing temperature.
In the other films $T_{anomaly}$ increases in a monotonic fashion with 
decreasing doping.
In the films \#Ox2 and \#Ox3 $T_{anomaly}$ represents rather a change 
of slope than a clear maximum.
Importantly, we find that $T_{anomaly}>T_c$ below optimal doping.
This is indicated in the inset, where in addition to $T_{anomaly}$ 
the parabolic dependence of the critical temperature according to 
Eq. (\ref{eqTc(p)}) with $T_{c,max}=90$ K is given.
Apparently, the origin of the varying coupling ratios is not a 
superconductivity-induced effect.
However, in the film \#Ox4 we find $T_{anomaly}=62\pm 6$ K which 
corresponds well to 
the temperature interval at which the $\rm B_{1g}$ $2\Delta$ peak passes 
the phonon as shown in Fig. \ref{figBackTemp}.
This, in turn, suggests a connection with the superconducting gap as 
proposed earlier.\cite{Altendorf93}

In the simple Fano fit according to Eq. (\ref{eqFano}) the asymmetry of 
the phonon is proportional to the inverse of the Fano parameter $q$.
Hence, a maximum value of the ratio depicted in Fig. \ref{figTanomaly} 
would correspond to a minimal value of the Fano parameter.
Using the simple Fano formula 
Altendorf {\em et al.}\cite{Altendorf93} studied the $\rm B_{1g}$ 
phonon in differently doped Y-123-O$\rm _{6+y}$ single crystals.
In agreement with our results they observe that the asymmetry $(1/q)$ 
exhibits a maximum as a function of temperature but could only resolve 
it in samples with oxygen contents $\mathrm{y}\geq 0.85$.
They find, that the temperature at which the maximum appears 
shifts to higher values with decreasing doping.
These temperatures correspond to the anomaly 
temperatures $T_{anomaly}$ defined in this study.
At a particular doping, the anomaly temperatures are higher in the 
crystals than in the films which might be correlated to the 
observation that the $\rm B_{1g}$ $2\Delta$ peaks of the crystals 
exceed those of the films at a particular doping level 
(see Fig. \ref{fig2DeltaDop}).

In addition to the anomalies with respect to intensity and coupling,
we noted in Sec. \ref{subsec:pse} that 
self-energy effects already started to appear above $T_c$
in all films whose doping levels are below the optimal value.
These self-energy effects cannot be related to background redistributions, 
especially, not in the underdoped films \#Ox1, \#Pr1, and \#Pr2 in which
$\rm B_{1g}$ $2\Delta$ peaks have not been observed.
The question arises whether the self-energy effects observed above $T_c$, 
i.e. softening and sharpening, 
can be an indication of a lattice distortion occurring at $T_{anomaly}$.
In this case we would expect to observe indications of the distortion 
in all other modes as well. 
As this is not the case in the films 
\#Ox1, \#Ox2, \#Pr2, and \#Pr3
we conclude that the self-energy effects as 
well as the anomalies with respect to coupling and intensity in the 
underdoped films
originate from the interaction with electronic excitations which are 
not directly observed in the Raman spectra.
Possible candidates for these excitations are those which are known to 
become strongly suppressed below the pseudogap temperature $T^*$ 
in underdoped cuprates.

Suppressions or gap-like structures have been observed with
NMR,\cite{Warren89} specific 
heat,\cite{Loram93} resistivity,\cite{Ito93}
optical conductivity,\cite{Homes93} ARPES,\cite{Norman98}
SIS tunneling,\cite{Miyakawa98} or
scanning tunneling microscopy.\cite{Renner98a}
They have also been reported in previous Raman studies in which the 
$\rm B_{1g}$ response of underdoped Bi-2212 compounds was 
investigated.\cite{Blumberg97,Quilty98}
In Y-123, however, reductions of spectral weight are only reported in 
$\rm B_{2g}$ symmetry which is not relevant here.\cite{Nemetschek97}
In agreement with the latter work and other 
studies\cite{Chen97,Hackl98}
we do not see significant change of the $\rm B_{1g}$ responses in the
underdoped RE-123 films below $T^*$.
Hence, we must conclude that these 
electronic excitations are not or only weakly Raman-active in the 
RE-123 system.

Recent ARPES experiments helped to clarify the $\mathbf{k}$ and 
$T$ dependence of the pseudogap.\cite{Harris96,Norman98}
The following picture emerges from these studies:
At $T=T^*$ a gap opens up 
at the FS crossings along the 
$(\frac{\pi}{a},0)$ to $(\frac{\pi}{a},\frac{\pi}{a})$ direction.
With decreasing temperature the gap value increases and 
spreads out over the FS. 
Simultaneously,
the non-gapped regions of the FS, so-called ``Fermi arcs'', shrink 
and collapse into the point node of the $d_{x^2-y^2}$-wave gap
at $T=T_c$.
The energy of the superconducting gap observed 
at the FS crossings below $T_c$ evolves 
smoothly from the values of the pseudogap observed above 
$T_c$ (see lowest panel in Fig. \ref{fig2DeltaTemp}).
No pseudogap features are observed in the normal state in overdoped 
samples.

Coming back to our observations one may assume that $T_{anomaly}$ is 
that temperature at which the
gap or the pseudogap is passing the phonon.
The doping dependence of $T_{anomaly}$ as seen in Fig. \ref{figTanomaly} 
yields strong support for this assumption, which we have already 
presented in a previous work.\cite{Bock99} 
In order to clarify this point let us consider the temperature 
dependence of the gap that we can read out of our data.
We will restrict ourselves to the energy of the maximum gap which 
appears near $(\frac{\pi}{a},0)$ as these regions are probed with 
preference in Raman experiments carried out in $\rm B_{1g}$ symmetry.
In fact, the $\rm B_{1g}$ $2\Delta$ peak has been identified with the 
maximum value of a $d_{x^2-y^2}$ gap in various 
theoretical treatments of the Raman 
response.\cite{Devereaux94,Krantz95,Jiang96,Strohm97}
As a measure of the temperature at which the gap closes 
around $(\frac{\pi}{a},0)$ 
we can use $T^*$ in conformity with the ARPES results.
In order to connect the $\rm B_{1g}$ $2\Delta$ peaks with the 
pseudogap temperature we will use a simple straight line which 
reflects the ARPES results to a good approximation.
The obtained graphs for the ordered films are depicted in 
Fig. \ref{fig2DeltaTemp}.
In the upper panel we show the results of the oxygen doped films which 
exhibited $2\Delta$ peaks.
The grey horizontal bar represents the energy of a typical 
renormalized $\rm B_{1g}$ phonon frequency of 342 cm$^{-1}$.
The open symbols on this bar represent the anomaly temperatures of the 
respective films.
In accordance with our assumption, they are located close to the intercepts of 
the linear extrapolations that connect the $2\Delta$ peaks observed 
near $T_c$ with the 
respective pseudogap temperatures in the films \#Ox2 and \#Ox3.
Interestingly, we observe $2\Delta$ peaks with energies below the 
phonon frequency in the film \#Ox4.
Here, we find that $T_{anomaly}$ lies right in that temperature range in 
which the $2\Delta$ peaks pass the phonon frequency.
In the middle panel the data of the ordered Pr- or Ca-doped films are 
depicted.
As mentioned earlier, the $2\Delta$ peaks in the film \#Ca2 are below 
the phonon frequency for all temperatures. This is in agreement with our 
inability to identify $T_{anomaly}$ in this film.
The data for the film \#Pr3 correspond well to the observations in 
the oxygen-doped samples.
As we did not observe any $2\Delta$ peaks in the film \#Pr2 only 
$T_{anomaly}$ and $T^*$ are given for this sample.

In the lowest panel of Fig. \ref{fig2DeltaTemp} leading edge shifts 
determined by Harris {\em et al.}\cite{Harris96} from an analysis 
of ARPES data are reproduced.
The ARPES spectra have been taken on an underdoped $T_c=85$ K 
Bi-2212 single crystal at the FS crossings along the 
$(\frac{\pi}{a},0)$ to $(\frac{\pi}{a},\frac{\pi}{a})$ direction.
They represent a measure of the gap value.
Apparently, these data closely resemble those of the underdoped 
$T_c=87$ K film \#Ox2 in the upper panel.
Regarding the different energy scales one has to take into account that 
ARPES measures $\Delta$ instead of $2\Delta$ and that the 
leading edge shift is given instead of the gap value.

The data depicted in Fig. \ref{fig2DeltaTemp} suggest 
strongly that the enhanced ratios $(g/g_{pp})^2$ observed 
at $T_{anomaly}$ have a quasi-resonant origin.
The appearance of a resonance in the bare phonon contribution
$I_{phon}=\pi A g_{pp}^2$ is in agreement with the assignment of
$g_{pp}$ to the vertex which couples the phonon to intraband 
electronic excitations given in Sec. \ref{subsec:mod}.
This abbreviated ``photon-phonon'' vertex explicitly allows resonances 
to occur as it involves $\mathbf{A}\cdot\mathbf{p}$ 
transitions.\cite{Hayes78}
A superconductivity-induced increase of the interband contribution 
to the Raman intensity of the $\rm B_{1g}$ phonon has been 
obtained in a theoretical treatment of the scattering 
process by Sherman {\em et al.}\cite{Sherman95}
The latter authors showed that the increase will depend on the doping 
level, becoming stronger in underdoped samples.
This appears to be in agreement with our data.
Our results, however, also indicate that there is an intimate relation
between the intensity anomaly due to the interband 
contribution and the relative energies of 
phonon and $\rm B_{1g}$ $2\Delta$ peak which has not been observed before. 
We find that the resonance in $I_{phon}$
vanishes as soon as the phonon energy is above $2\Delta(T=0)$.

In IR Raman measurements of underdoped Y-123 with $T_c\leq 84$ K,
an anomalously increasing intensity of a broad low-energy electronic feature
has been observed blow $T^*$.\cite{Ruani97}
The feature is assigned to an interband excitation which involves
$\mathbf{A}\cdot\mathbf{p}$ transitions.
On the basis of our results we predict that the anomaly will vanish at 
higher dopings when $2\Delta < 500$ cm$^{-1}$.

The resonance with respect to the effective coupling could have two 
different origins.
First, it can simply occur because the background contribution at the 
phonon frequency becomes maximal when the gap value equals the phonon 
energy, or second, it may indicate that the electron-phonon coupling 
itself becomes resonantly enhanced.
The first can only be expected as long as $2\Delta$ peaks are 
observed. 
In the underdoped films, however, we observe a structureless response 
which is not altered within the noise level for all temperatures.
The second argument would indicate a breakdown of Migdal's 
theorem which allows one to neglect vertex corrections to
the electron-phonon coupling in a metal.\cite{Migdal}
Whether cuprate HTS are a class of materials where such a breakdown 
could occur is currently under debate\cite{Migdal} in view of recent 
experiments which indicate that polaronic charge carriers are present 
in superconducting $\rm La_{2-x}Sr_xCuO_4$.\cite{Mueller} 
That the $\rm B_{1g}$ phonon interacts with the pseudogap in the 
observed way is most likely related to the fact that the 
electron-phonon vertex, which has been obtained in a first 
approximation by Devereaux et al,\cite{Devereaux95Pho}
samples the FS with preference along the $k_x$ and $k_y$ directions.
Hence, it will be most sensitive to the maximum value of the gap 
which appears in these regions.

As a $\rm B_{1g}$ plane oxygen phonon is present in all cuprates with 
more than one CuO$_2$ plane in the unit cell, we predict that phonon 
anomalies will also be observed in these compounds when the $2\Delta$ 
peak merges the phonon frequency.
The effect will be smaller in those compounds in which the planes are 
not buckled.

\section{conclusion}
\label{sec:con}
We have investigated laser-deposited $c$-axis oriented
$\rm Y_{1-x}(Pr,Ca)_xBa_2Cu_3O_{6+y}$ thin films using 
Raman and transport measurements, and the Raman spectra of an 
$\rm YBa_2Cu_3O_{7}$  single crystal.
Variations of the Pr, the Ca, or the chain-oxygen content allow us 
to achieve different doping levels ranging from under- to overdoped.
Using a previously presented description of the Raman response we are 
able to deconvolute electronic and phononic excitations.
This allows us to identify the $2\Delta$ peaks appearing below $T_c$
even in the presence of strongly interacting phonons.
Whereas we observe that the energy of the $\rm A_{1g}$ 
$2\Delta$ peak follows $T_c$, the peak energies in $\rm B_{1g}$ 
symmetry increase monotonically with decreasing doping.
As the $\rm B_{1g}$ $2\Delta$ peaks are measures of the maximum value 
of a $d_{x^2-y^2}$-wave gap, the latter result is in qualitative
agreement with the trends obtained in SIS-tunneling 
or ARPES studies of Bi-2212 single crystals
where this non-mean field behavior has already been established. 
At low dopings the $\rm B_{1g}$ $2\Delta$ peaks become unresolvable.
The $\rm A_{1g}$ $2\Delta$ peaks, however, are still observed
at low dopings in Pr-doped films in contrast to previous results 
obtained in oxygen-reduced samples.
We relate this finding to persisting chain-related bands which inhibit 
a screening of the $\rm A_{1g}$ response expected in one-band models 
of the Raman response.
In Pr-doped films
we find enhanced scattering intensities for all frequencies in $\rm 
A_{1g}$ as well as in $\rm B_{1g}$ symmetry compared to the oxygen- 
or Ca-doped films.
We attribute this observation to quasi-particles scattering at the 
Pr atoms where Pr $4f$ orbitals have hybridized with adjacent oxygen $2p$ 
orbitals into the Fehrenbacher-Rice state.
In disordered Pr- and Ca-doped films we observe weakenings of the 
superconductivity-induced features accompanied by decreased dopings.

Regarding phononic excitations we focus in this work on the $\rm B_{1g}$ 
phonon at 340 cm$^{-1}$, a mode which involves out-of-phase vibrations 
of the plane-oxygen atoms.
This mode interacts strongly with low-lying electronic excitations and 
hence, exhibits an asymmetric Fano-type line shape.
Our description of the $\rm B_{1g}$ phonon enables us to identify the 
self-energy contributions due to the electronic response and, 
especially, yields measures of the effective electron-phonon coupling.
One of these measures is the mass-enhancement factor $\lambda$
which is found to agree well with LDA expectations for 
samples near optimal doping. 
Whereas the observed self-energy effects are in qualitative agreement 
with the relative positions of $\rm B_{1g}$ $2\Delta$ peak and phonon 
frequency, they cannot completely be attributed to the redistribution 
of the electronic response appearing below $T_c$.
Moreover, renormalizations above $T_c$, which we have observed below 
optimal doping, strongly indicate that the phonon also 
interacts with excitations which become gapped in the pseudogap phase.
It turns out that these excitations are not directly observed in the 
Raman spectra.
However, we find that the temperature at which the pseudogap passes the 
phonon can be identified as that one at which the phonon is maximal 
asymmetric.
Using our model we find that the ratio of the 
electron-phonon coupling constants to intra- and interband electronic 
excitations is at maximum at that temperature.
Pseudogap temperatures $T^*$ determined from transport 
measurements, temperatures of maximum asymmetry, and 
$\rm B_{1g}$ $2\Delta$ peaks yield temperature dependent gaps for our 
films which closely resemble those seen in ARPES measurements 
at the FS crossings along the
$(\frac{\pi}{a},0)$ to $(\frac{\pi}{a},\frac{\pi}{a})$ direction.

\acknowledgements

The author thanks G. Blumberg, M. K\"all, D. Manske, S. Ostertun, 
and M. R\"ubhausen for careful reading of the manuscript. 
Special thanks are expressed to M. K\"all for providing the single 
crystal data and to R. Das Sharma, S. Ostertun and M. R\"ubhausen for 
the cooperation.
Intensive discussions with M.V. Klein, U. Merkt, C.T. Rieck, 
and E.Y. Sherman are highly appreciated.
Parts of this work have been supported by the German Science 
Foundation via the Graduiertenkolleg 
``Physik nanostrukturierter Festk\"orper'' and by the 
Bundesministerium f\"ur Bildung, Forschung, Wissenschaft und 
Technologie, Germany, under contract number 13N6734/0.

\newpage
\begin{figure}
\caption{Temperature dependence of the in-plane resistivity of the 
   investigated
   $\rm Y_{1-x}(Pr,Ca)_xBa_2Cu_3O_{6+y}$ films.}
\label{figRes}
\end{figure}
\begin{figure}
\caption{Illustration of the fit parameters.
    In the left panel on the top an artificial efficiency 
    consisting of an interacting phonon as well as of an electronic 
    response is shown. Below, the interacting phonon contribution,
    proportional to $(1/C)$, with 
    renormalized parameters is displayed. 
    At the bottom, the non-interacting phonon contribution,
    proportional to $R_0^2/C$, with bare 
    phonon parameters is given. 
    The electronic response used to obtain the 
    interacting phonon contribution is shown in the top row of the 
    right panel. Below, the two Lorentzians describing the 
    redistribution of the background are displayed.
    The $2\Delta$ peak is described with the frequency $\omega_{2\Delta}$ 
    and the linewidth (HWHM) $\Gamma_{2\Delta}$.
    The intensities of the peak and the suppression are  
    $I_{2\Delta}$ and $I_{supp}$, respectively.
    At the bottom, the hyperbolic tangent with crossover frequency 
    $\omega_T$ and intensity $I_{\infty}$ is shown.
    The offsets of the contributions are indicated by horizontal lines.}
\label{figVis}
\end{figure}
\begin{figure}
\caption{Raman efficiencies of the film \#Ox4 in $\rm A_{1g}$, $\rm B_{1g}$, 
  and $\rm B_{2g}$ symmetry taken at
  $T=152$ K (solid lines) and $18$ K (dots). 
  Raman-active phonons are indicated.
  The efficiencies are offset as given in brackets. }
\label{figAllGeom}
\end{figure}
\begin{figure}
\caption{(a): $\rm A_{1g}$ symmetry.
  In the uppermost spectrum the Raman efficiency $I_0(\omega)$ 
  of the film \#Ox4 taken at 
  $T=18$ K (dots) and the fit result (solid line) are shown.
  Below, the fitted phonon profiles (solid line),
  the electronic response $\varrho_*(\omega)$ obtained after 
  subtraction of the phonons (dots), and the numerically determined 
  real part of the electronic response function $R_{*}(\omega)$ (dots) 
  are given.
  The analytical descriptions of the response used in the fit 
  are included as solid lines. 
  The spectra are offset as given in brackets, all intensities are given 
  in the same units. 
  (b): The corresponding data in $\rm B_{1g}$ symmetry.}
\label{figFit}
\end{figure}
\begin{figure}
\caption{Electronic responses $\varrho_*(\omega)$ of the film \#Ox4 
  in $\rm A_{1g}$ and 
  $\rm B_{1g}$ symmetry at various temperatures.  
  The dashed lines indicate the 
  energies of the fit parameter $\omega_{2\Delta}$ at $T=18$ K. 
  In the right panel the frequency of the $\rm B_{1g}$ 
  phonon is given as a dashed line.
  The spectra are offset for clarity.
  }
\label{figBackTemp}
\end{figure}
\begin{figure}
\caption{Electronic responses $\varrho_*(\omega)$
  of the oxygen-doped films in $\rm A_{1g}$ and 
  $\rm B_{1g}$ symmetry at $T=18$ K. 
  The dashed lines indicate the energies of the fit parameter 
  $\omega_{2\Delta}$ of the 
  film \#Ox4 in the respective symmetries at $T=18$ K.
  The spectra are offset as given in brackets.
  }
\label{figBackOx}
\end{figure}
\begin{figure}
\caption{Temperature dependence of fit parameters of the $\rm B_{1g}$ 
  phonon for the oxygen-doped films and the single crystal.
  Doping increases from left to right.
  Closed circles represent the bare phonon parameters 
  $\omega_p$, $\Gamma$, and $I_{phon}$, 
  open circles the renormalized values 
  $\omega_{\nu}$, $\gamma_p$ and $I_{tot}$.
  Diamonds and crosses are the self-energy contributions 
  $R_*(\omega_p)/C $ and $\varrho_*(\omega_p)/ C$, respectively.
  Dashed lines are fits to an anharmonic decay, solid ones are guides 
  to the eye. 
  Also shown is the ratio $I_{\infty}/C$ (closed squares).
  Dash-dotted lines indicate the respective $T_c$'s of 
  the films and the single crystal.
  Some data are scaled as given in brackets.
  Marker sizes represent the vertical accuracies.}
\label{figPhonOx}
\end{figure}
\begin{figure}
\caption{(a): Raman efficiencies $I_0(\omega)$ of the 
  single crystal in $\rm A_{1g}+B_{2g}$ [$z(x'x')\overline{z}$]' and 
  $\rm B_{1g}$ symmetry taken at $T=12$ K.
  (b): Electronic responses $\varrho_*(\omega)$ obtained after 
  subtraction of the phonons.}
\label{figSpecCrys}
\end{figure}
\begin{figure}
\caption{Electronic responses $\varrho_*(\omega)$ 
  of the ordered substituted films in $\rm A_{1g}$ and 
  $\rm B_{1g}$ symmetry at $T=18$ K. 
  The dashed lines indicate the energies of the fit parameter 
  $\omega_{2\Delta}$ of the 
  film \#Ox4 in the respective symmetries at $T=18$ K.
  The spectra are offset as given in brackets.
  }
\label{figBackOrd}
\end{figure}
\begin{figure}
\caption{Temperature dependence of the fit parameters of the $\rm B_{1g}$ 
  phonon for the substituted films.
  Doping increases from left to right.
  Columns of the disordered films have a grey background.
  Closed circles represent the bare phonon parameters 
  $\omega_p$, $\Gamma$, and $I_{phon}$, 
  open circles the renormalized values 
  $\omega_{\nu}$, $\gamma_p$ and $I_{tot}$.
  Diamonds and crosses are the self-energy contributions 
  $R_*(\omega_p)/C $ and $\varrho_*(\omega_p)/ C$, respectively.
  Dashed lines are fits to an anharmonic decay, solid ones are guides 
  to the eye. 
  Also shown is the ratio $I_{\infty}/C$ (closed squares).
  Dash-dotted lines indicate the respective $T_c$'s of 
  the films and the single crystal.
  Some data are scaled as given in brackets.
  Marker sizes represent the vertical accuracies}
\label{figPhonSubst}
\end{figure}
\begin{figure}
\caption{Electronic responses $\varrho_*(\omega)$ 
   of the order-disorder pairs with 20 \% Pr and 
  5 \% Ca content in $\rm A_{1g}$ 
  (dotted line) and $\rm B_{1g}$ (solid line) symmetry at $T=18$ K. 
  All four panels have the same frequency and intensity range.
  }
\label{figBackDis}
\end{figure}
\begin{figure}
\caption{Normalized resistivities $[\rho(T)-\rho(O)]/\alpha T$ as a 
  function of temperature, where $\alpha$ is the slope of the $T$-linear 
  region of $\rho(T)$ and $\rho(0)$ is the extrapolated value to $T=0$ K.
  }
\label{figPseudo}
\end{figure}
\begin{figure}
\caption{Self-energy effects of the 
 investigated films (solid circles) as well as of single crystals 
 (open circles) versus energy of the $\rm B_{1g}$ $2\Delta$ peak.
 In the upper (lower) panel the broadening (frequency shift) relative 
 to the anharmonic decay for the data points at the lowest 
 temperatures are given.
 Results of studies in which the $2\Delta$ peaks and 
 the self-energy effects of identical single crystals have been 
 obtained separately by 
 Chen {\em et al.}\protect\cite{Chen93} 
 and Altendorf {\em et al.}\protect\cite{Altendorf93}, 
 respectively, are included.
 The dotted lines are guides to the eye and the grey vertical bar 
 indicates a typical phonon frequency of 
 $\omega_{\nu}=342$ cm$^{-1}$.
 The data of the disordered film \#Ca1 are given in brackets.}
\label{figDopSE}
\end{figure}
\begin{figure}
\caption{ Doping dependence of the $2\Delta$ peaks 
  of the investigated films (solid circles) as well as of single 
  crystals (open circles) in $\rm A_{1g}$ and $\rm B_{1g}$ 
  symmetry at $T= 18$ K.
  Single crystal data of Chen {\em et al.}\protect\cite{Chen93} 
  are also included.
  Solid lines represent a standard BCS mean-field $d$-wave 
  prediction\protect\cite{Won94} with $2\Delta=4.28$ 
  $\mathrm{k}_B$$T_c$, shown to 
  highlight the non-mean-field trend of the $\rm B_{1g}$ data.
  The dashed line describes the doping dependent gap $2\Delta$ in 
  Bi-2212 obtained from an analysis of ARPES and tunneling 
  in Ref.\protect\onlinecite{Miyakawa98}. 
  }
\label{fig2DeltaDop}
\end{figure}
\begin{figure}
\caption{Doping dependence of the electronic responses 
  $\varrho_*(\omega)$ 
  of the underdoped films \#Pr2 and \#Ox1, the optimally doped 
  film \#Ox3, and the overdoped film \#Ca2 
  in $\rm A_{1g}$ and $\rm B_{1g}$ symmetry at $T=18$ K.
  In order to reduce the noise, 10 neighboring data points of the 
  original spectra have been averaged for this graph yielding a 
  spectral resolution of $\approx 20$ cm$^{-1}$.
  }
\label{figBackDop}
\end{figure}
\begin{figure}
\caption{(a): Comparison of the $\rm B_{2g}$ Raman efficiencies 
  $I_0(\omega)$ of the films \#Pr2 and \#Ox3 taken at 18 K.
  (b) Electronic response $\varrho_*(\omega)$
  in $\rm B_{2g}$ symmetry of the film \#Ca2 at 18 K (solid line) and 
  102 K (dotted line).
  The dashed line indicates the energy of the fit parameter 
  $\omega_{2\Delta}$ at $T=18$ K.
  }
\label{figBackB2g}
\end{figure}
\begin{figure}
\caption{Doping dependence of the Raman efficiencies $I_0(\omega)$ 
  in $\rm B_{1g}$ symmetry taken at $T=18$ K.
  Arrows indicate phonons of the substrate which remained after 
  subtraction of the substrate signal.
  For each sample the respective effective coupling constant $I_{\infty}/C$
  is given. 
 The dashed line indicates a typical phonon frequency of 342 
 cm$^{-1}$.
 The spectra are offset as given in brackets.
 }
\label{figBackB1g}
\end{figure}
\begin{figure}
\caption{Normalized ratio of the square of the
  coupling constants $g$ and $g_{pp}$ versus temperature.
  Doping decreases from top to bottom and
  open circles and crosses are used alternately for clarity.
  Left panel: oxygen-doped films and single crystal, right panel: 
  substituted films. 
  Fat bars indicate temperatures $T_{anomaly}$ at which the 
  ratio is either at maximum or changes slope.
  Dotted lines represents the maximum transition 
  temperature of the films $T_{c,max}=90$ K.
  Data are offset as given in brackets.
  In the inset $T_{anomaly}$ is given versus doping. 
  The open square corresponds to the single crystal and closed 
  ones to the films. 
  The data point of the
  film \#Ca2 is given as an extended rectangle.
  Also shown in the inset is a $T_c(p)$ curve obtained according to 
  Eq. (\protect\ref{eqTc(p)}) with $T_{c,max}=90$ K.
  }
  \label{figTanomaly}
\end{figure}
%
\begin{figure}
\caption{Energy of the gap $2\Delta$ near $(\frac{\pi}{a},0)$ 
  determined by the fit parameters $\omega_{2\Delta}$ in 
  $\rm B_{1g}$ symmetry and $T^*$ (closed symbols)
  versus temperature.
  A zero gap energy is indicated by a dotted line and a gap energy of 
  342 cm$^{-1}$ by a grey bar. 
  Open symbols on the grey bar represent $T_{anomaly}$
  of the respective films.
  In the lower panel ARPES results of the shifts of the leading edge 
  midpoints relative to the Fermi energy at the FS crossing near 
  $(\frac{\pi}{a}, 0.2\frac{\pi}{a})$ taken form Ref. 
  \protect\onlinecite{Harris96} are depicted.
  These data have been obtained on an underdoped Bi-2212 single 
  crystal and represent a continuously decreasing gap.
  }
  \label{fig2DeltaTemp}
\end{figure}

\newpage %

\widetext

\begin{table}
  \caption{Names and properties of the investigated
  $\rm Y_{1-x}(Pr,Ca)_xBa_2Cu_3O_{6+y}$ films
  with nominal Ca or Pr contents $\mathrm{x}$ and cool down 
  oxygen-partial pressures $p_{ox}$.
  The film thicknesses are determined from the deposition 
  parameters.\protect\cite{Heinsohn98}
  All films are deposited on 1 mm thick SrTiO$_3$(100) 
  substrates.
  $\rho$, $T_{c}$, and $T^*$ are the resistivity, the critical 
  temperature, and the pseudogap temperature, respectively.
  $p$ is the doping level and values 
  determined from the Ca, Pr, and oxygen content as well 
  as values determined 
  according to Eq. (\protect\ref{eqTc(p)}) are given.
  In this study the mean value $\bar{p}$ is used.
   }
\label{tabProp}
  \begin{tabular}{lcccccccccc}

Film&Thickness&x &$p_{ox}$  &y
       &$\rho(100\,K)$            
       &$T_{c}$&$T^*$&$p(x;y)$&$p(T_c;T_{c,max})$&$\bar{p}$\\
       &(nm)&(\%)  &(mbar) &            
       &($\rm \mu\Omega$cm)&(K)          &(K)     &             &      &\\
\hline
\#Pr1 &240&Pr: 20&1000 &$1.0\pm0.05$
       &311&61.2&$92\pm 4$&($0.114\pm0.005$)&$0.100\pm0.002$&
       $0.100\pm0.005$\\
\#Pr2 &240&Pr: 20&1000 &$1.0\pm0.05$
       &221&71.5&$155\pm5$&$0.114\pm0.005$&$0.112\pm0.002$&
       $0.112\pm0.003$\\
\#Pr3 &324&Pr: 10&1000 &$1.0\pm0.05$
       &73&86.3&$130\pm 13$&$0.151\pm0.01$&$0.145\pm0.005$&
      $0.147\pm0.008$\\
\#Ox1 &270&0&1 &$0.52\pm0.03$
       &348&56.0&$(255\pm 5)$&$0.086\pm0.006$&$0.093\pm0.001$&
       $0.092\pm0.002$\\
\#Ox2 &324&0&3.5 &$0.85\pm0.05$
       &162&87.0&$208\pm 7$&$0.156\pm0.01$&$0.140\pm0.005$&
       $0.145\pm0.008$\\
\#Ox3 &324&0&50 &$0.92\pm0.05$ 
       &119&89.8&$155\pm 5$&$0.170\pm0.01$&$0.160\pm0.005$&
       $0.163\pm0.008$\\
\#Ox4 &270&0&1000 &$1.0\pm0.05$
       &86&88.0&$102\pm 3$&$0.187\pm0.01$&$0.176\pm0.005$&
       $0.180\pm0.008$\\
\#Ca1 &324&Ca: 5&1000 &$1.0\pm0.05$
       &60&85.0&$133\pm 12$&($0.212\pm0.01$)&$0.186\pm0.005$&
       $0.186\pm0.01$\\
\#Ca2 &324&Ca: 5&1000 &$1.0\pm0.05$
       &97&82.7&$93\pm1$&$0.212\pm0.01$&$0.191\pm0.005$&
       $0.198\pm0.008$
  \end{tabular}
\end{table}  
\narrowtext

\begin{table}
\caption{Analyses of the electronic responses and 
  the self-energy effects of the $\rm B_{1g}$ phonon  at $T=18$ K of 
  the $\rm Y_{1-x}(Pr,Ca)_xBa_2Cu_3O_{6+y}$ films and the single 
  crystal.
  The energies $2\Delta \mathrm{(\sigma)}$ represent the center 
  frequencies of the peaks in the electronic responses in $\sigma$ 
  symmetry ($2\Delta$ peaks).
  Also given are the exponents $z$ that describe the slopes $\omega^z$ 
  of the electronic responses for $\omega \rightarrow 0$.
  The values $\Delta\omega_{\nu}$ and $\Delta\gamma_p$ 
  describe the frequency shifts and broadenings relative 
  to the anharmonic decay.
  They are given in brackets for the film \#Pr1 in which a lattice 
  distortion is observed around 150 K.
  }
\label{tabAnal}
  \begin{tabular}{lcccccc}
    
Film&$2\Delta \mathrm{(A_{1g})}$  &$2\Delta \mathrm{(B_{1g})}$
       &$\omega^z \mathrm{(A_{1g})}$&$\omega^z \mathrm{(B_{1g})}$
       &$\Delta\omega_{\nu}$&$\Delta\gamma_p$   \\   
       &(cm$^{-1}$)&(cm$^{-1}$) & &            
       &(cm$^{-1}$)&(cm$^{-1}$)  \\
\hline
\#Pr1 &$270\pm 20$&--                 &0.5&-- &(-4.6)&(-1.0)  \\
\#Pr2 &$290\pm 20$&--                 &0.9&0.4&-1.6&-2.5  \\
\#Pr3 &$325\pm 10$&$580\pm 10$&1.2&0.6&-4.9&-1.5  \\
\#Ox1 &--                 &--                 &-- &0.2&-0.3&-0.3  \\
\#Ox2 &$295\pm 10$&$540\pm 25$&0.8&0.4&-3.5&-0.8  \\
\#Ox3 &$305\pm 10$&$515\pm 20$&1.2&0.9&-5.6&-0.5  \\
\#Ox4 &$280\pm 20$&$390\pm 10$&1.0&1.0&-6.7&5.3  \\
\#Ca1 &$270\pm 20$&$350\pm 20$&0.9&0.5&-4.0&3.1  \\
\#Ca2 &$245\pm 20$&$305\pm 20$&1.2&1.7&0.8 &5.4  \\
\#sc   &$335\pm 15$&$435\pm 10$&1.1&1.0&-8.6&1.9 
  \end{tabular}
\end{table}  


\begin{references}
\bibitem{KKRvonTanH} The Hilbert transformed incoherent 
contribution is: \\ 
   $R_{inc}(\omega)= 
   I_{\infty}\left[0.5 + \frac{2}{\pi}\, {\rm ln}(\omega_{cut}/\omega_T) +
                           \sqrt{2} \, {\rm artanh}(\eta-1)\right]$, with
  $\eta=\tanh(\omega/\omega_T)/(\omega/\omega_T)$.
\bibitem{Int-230(y)} The probability to have an interrupted chain at 
a particular oxygen content $\mathrm{y}$ can be described by:
$P(\mathrm{y})\approx 0.484\mathrm{y}-5.285\mathrm{y}^2+
79.391\mathrm{y}^3-177.29\mathrm{y}^4 +
138.85\mathrm{y}^5-36.131\mathrm{y}^6$
for $0\leq \mathrm{y} \leq 1$.
\end{references}
\end{document}